\documentclass[
aip,
amsmath,amssymb,
preprint
]{revtex4-1}

\usepackage{graphicx}
\usepackage{dcolumn}
\usepackage{bm}

\usepackage[utf8]{inputenc}
\usepackage[T1]{fontenc}

\usepackage{footnoterange}
\usepackage{amsmath}
\usepackage{amsthm}
\usepackage{amssymb}
\usepackage{mathrsfs}
\usepackage{mathtools}
\usepackage{epsfig}
\usepackage{caption}
\usepackage{subcaption}

\usepackage{color}
\usepackage{wasysym}
\usepackage{booktabs}
\usepackage{multirow}
\usepackage{siunitx}
\usepackage{natbib}

\usepackage{epstopdf}
\usepackage[dvipsnames]{xcolor}
\usepackage{tikz}
\usepackage{tabularx}
\usepackage{ragged2e}

\usepackage{pifont}
\usepackage{xcolor}
\usepackage{hhline}

 \usepackage[colorlinks,
            colorlinks=true,
            linkcolor=blue,
            citecolor=blue,
            filecolor=magenta,
            urlcolor=black]{hyperref}
               
\begin{document}

\preprint{AIP/123-QED}

\title{Anomalous Features in Internal Cylinder Flow Instabilities subject to Uncertain Rotational Effects}

\author{Ali Akhavan-Safaei}
\affiliation{Department of Mechanical Engineering, Michigan State University, 428 S Shaw Ln, East Lansing, MI 48824, USA}
\affiliation{Department of Computational Mathematics, Science, and Engineering, Michigan State University, 428 S Shaw Ln, East Lansing, MI 48824, USA}
\author{S. Hadi Seyedi}
\affiliation{Department of Mechanical Engineering, Michigan State University, 428 S Shaw Ln, East Lansing, MI 48824, USA}
\author{Mohsen Zayernouri} \email{zayern@msu.edu}
\affiliation{Department of Mechanical Engineering, Michigan State University, 428 S Shaw Ln, East Lansing, MI 48824, USA}
\affiliation{Department of Statistics and Probability, Michigan State University, 619 Red Cedar Road, Wells Hall, East Lansing, MI 48824, USA}

\date{\today}

\begin{abstract}
We study the flow dynamics inside a high-speed rotating cylinder after introducing strong symmetry-breaking disturbance factors at cylinder wall motion. We propose and formulate a mathematically robust stochastic model for the rotational motion of cylinder wall alongside the stochastic representation of incompressible Navier-Stokes equations. We employ a comprehensive stochastic computational fluid dynamics framework combining spectral/$hp$ element method and probabilistic collocation method to obtain high-fidelity realizations of our mathematical model in order to quantify the propagation of parametric uncertainty for dynamics-representative quantities of interests. We observe that the modeled symmetry-breaking disturbances cause a flow instability arising from the wall. Utilizing global sensitivity analysis approaches, we identify the dominant source of uncertainty in our proposed model. We next perform a qualitative and quantitative statistical analysis on the fluctuating fields characterizing the fingerprints and measures of intense and rapidly evolving non-Gaussian behavior through space and time. We claim that such non-Gaussian statistics essentially emerge and evolve due to an intensified presence of coherent vortical motions initially triggered by the flow instability due to symmetry-breaking rotation of the cylinder. We show that this mechanism causes memory effects in the flow dynamics in a way that noticeable anomaly in the time-scaling of enstrophy record is observed in the long run apart from the onset of instability. Our findings suggest an effective strategy to exploit controlled flow instabilities in order to enhance the turbulent mixing in engineering applications.
\end{abstract}

\maketitle

\section{Introduction}\label{sec: Intro}

Understanding, quantifying, and exploiting \textit{anomalous transport} opens up a rich field, which can transform our perspective towards the extraordinary processes in thermo-fluid problems. This emerging class of physical phenomena refers to fascinating and realistic processes that exhibit non-Markovian (long-range memory) effects, non-Fickian (nonlocal) interactions, non-ergodic statistics, and non-equilibrium dynamics \cite{klages2008anomalous}. It is observed in a wide variety of complex, multi-scale, and multi-physics systems such as: sub-/super-diffusion in brain, kinetic plasma turbulence, aging of polymers, glassy materials, amorphous semiconductors, biological cells, heterogeneous tissues, and disordered media. 

Of particular interest, the structure of chaotic and turbulent flows is in a way that nonlocal and memory effects cannot be ruled out \cite{davidson2015turbulence, egolf2019nonlinear}. In fact, anomalous transport can essentially manifest in heavy-tailed and asymmetric distributions, sharp peaks, jumps, and self-similarities in the time-series data of fluctuating velocity/vorticity fields. Flow within and around cylinders is a rich physical problem that involves complex geometry and nonlinear flow instabilities, with unsolved questions on flow/vortex structures and anomalous turbulent mixing \cite{avila2011onset}. Numerous researchers have studied the flow and heat transfer characteristics when a fluid flow encounters a cylinder. These studies include fixed, cross-flow oscillations, inline oscillations, and rotation of the cylinder cases. Studies related to the interactions of the flow and moving bodies were first conducted by Strouhal in 1878. Gerrard \cite{gerrard1966mechanics} proposed a model for the vortex shedding mechanism and the resulted von K\'{a}r\'{a}mn vortex street. Effects of cross-flow and inline oscillations of a cylinder on vortex shedding frequency were first determined by Koopman \cite{koopmann1967vortex} and Griffin and Ramberg \cite{griffin1976vortex}, respectively. These studies are categorized as external flows around cylinders and some significant contributions in this regard may be found in \cite{barbi1986vortex, ongoren1988flow, ongoren1988flow2, dennis2000flow, nobari2006numerical}. However, flow inside systems with fast rotation including cylinders, squares and annulus geometries are also of great importance. Turbo-machinery, mixing process, gravity-based separators, geophysical flows, and journal bearing lubrication are all clear examples for these types of internal flows \cite{del2005stability, lappa2012rotating}.

 Moreover, in rotational cylinder flows, the flow may face a concave wall and centrifugal instabilities may be developed when the thickness of boundary layer is comparable to the radius of the curvature. Consequently, centrifugal instabilities lead to formation of stream-wise oriented vortices that commonly called Taylor-G\"{o}rtler vortices. These vortices can change the flow regime through a transition process to turbulence \cite{Kim2006, Sauret2012_PoF, kaiser2020}. In particular, the Taylor problem in Couette flow between two concentric rotating cylinders is another well-known example of centrifugal instabilities in rotating systems, which have been studied experimentally \cite{Gollub_1975, Dubrulle2005_PoF, Yufeng2014_PoF} and numerically \cite{Lagrange2008, ostilla_2013_DNS, Tang2015_DNS, Crowley_2020}. In such problems, emergence of the adverse angular momentum is an important mechanisms, which initiates flow instability. More specifically, Lopez \textit{et al.} \cite{lopez_2002_JFM} studied flow in a fully-filled rotating cylinder, which is driven by the counter-rotation of the endwall and found out that in the presence of considerably large counter-rotation, the separation of the Ekman layer from the endwall generates an unstable free shear layer that separates flow regions against the azimuth velocity. In fact, this shear layer is highly sensitive to the sources of disturbance appearing in the azimuth velocity, which essentially breaks the symmetry in the flow. Other symmetry-breaking effects were further investigated when they are originated from other sources such as inertial waves \cite{lopez2011_JFM}, oscillating sidewalls \cite{Lopez2010_PoF} and, precessional forcing \cite{lopez_2018}. 
 
 Inspired by the flow dynamics after the emergence of symmetry-breaking factors, we are specifically interested in computational study of the onset of flow instabilities and their long-time effects. To model such symmetry-breaking effects in rotational motion of cylinder, we introduce some featured sources of disturbance in angular velocity, which may be coupled by eccentricity rotation of the system. In reality, these sorts of symmetry-breaking noises could be a direct result of unexpected failure in the electro-mechanical rotational system/fixture, which may be accompanied by secondary inertial disturbances that intensify the instability and transition of the flow regime. From a mathematical modeling and simulation point of view, a deterministic view would inevitably fail to reflect the true physics of such highly complex phenomenon, which is involved with numerous sources of stochasticity, (\textit{i.e.}, sources of disturbance). This urges for another level of modeling and investigation, which respects the random nature of the problem and is capable of addressing the effects of such sources of randomness in the response of system. In general, these sources of randomness could be categorized into either \textit{aleatory} or \textit{epistemic} model uncertainties. Aleatory uncertainty affects the quantities of interest (QoI) by the natural variations of the model inputs and usually are hard to be reduced; nevertheless, epistemic uncertainty mostly comes from our limited knowledge on what we are modeling and could be stochastically modeled once we obtain additional information about the system \cite{KIUREGHIAN2009}. Uncertainty in modeling procedure and also inaccuracy of the measured data are two main factors in arising epistemic uncertainty. The uncertainty in modeling could be the result of a variety of possibilities including the effects of geometry \cite{TARTAKOVSKY2006, ZAYERNOURI2013_roughness, Berrone2018, asgharzadeh2019non, KWON2020_roughness}, constitutive laws \cite{HAMDIA2015, VUBAC2015, Prudencio2015, FARRELL2015, Morrison2018, Faghihi2018, seyedi2018multiresolution, seyedi2019high, Varghaei2019, suzuki2019thermodynamically, DeMoraes2020, DeMoraes2020data, yangmethod, ambartsumyan2020stochastic}, rheological models \cite{AkhavanSafaei_2018, Patra2018comparative, Patra2018}, low-fidelity and reduced-order modeling \cite{xiu2007parametric, gorle2013framework, KHALIL2015, gorle2019epistemic, NADIGA2019, Babaee2019, XIAO2019, mishra2019estimating, CORTESI2020, lu2020lagrangian, HAO2020_RANS-UQ1, HAO2020_RANS-UQ2, zhao2020reduced}, and random forcing sources in addition to the random field boundary/initial conditions \cite{xiu2003modeling, xiu2005high, Babee2013_a, Babee2013_b, khalil2014independent, BABAEE2017, Kharazmi2020}. In the current work, we seek to fill a gap in the rich literature of investigating flow instabilities inside rotating flow systems by emphasizing on the stochastic modeling of the fluid dynamics and later focusing on the anomalies in the anomalous transport features of such system through statistical and scaling analysis of the response. This goal is achieved through a comprehensive computational framework that employs high-fidelity flow simulator as ``forward solver'' in our stochastic model. Our forward solver employs a two-dimensional (2-D) computational model for rotating cylinder that is assumed to be fully-filled with the Newtonian fluid and the entire system is in rigid-body rotation state with the angular velocity of $\dot{\theta}_0 = d\theta/dt\vert_{t=0}$. In fact, this solid-body rotation state is a stable flow regime (with perfect rotational symmetry) that we take as the initial modeling stage where we introduce the symmetry-breaking disturbances in terms of ``oscillatory'' and ``decaying'' angular velocity for the cylinder's wall. Such angular velocity model in addition to the effects of ``eccentric rotation'' would make a strong symmetry-breaking effect (disturbance model) to study the dynamics of instability while our model addresses the stochastic nature of the problem. The main contributions of our study are highlighted in the following items:
 \begin{itemize}
     \item We formulate stochastic Navier-Stokes equations subject to random symmetry-breaking inputs, affecting the incompressible flow within a high-speed rotating cylinder. We employ spectral element method (SEM) along with the probabilistic collocation method (PCM) to formulate a stochastic computational framework.
     \item We perform a global sensitivity analysis and reduce the dimension of random space to the dominant stochastic directions. We compute the expected velocity field enabling us to obtain the fluctuating part of the velocity at the onset of flow instabilities induced by the modeled symmetry-breaking effects. Computing the velocity fluctuations lets us study the temporal evolution of their probability distribution function, which sheds light on the instability dynamics and anomalous transport features.
     \item Obtaining the fluctuating vorticity field, we identify a well-pronounced and evolving non-Gaussian statistical behavior at the onset of flow instability essentially implying that the disturbances (influencing the cylinder rotation) cause generation of ``coherent vortical structures''. These vortices increase the memory effects in the hydrodynamics and we characterize their impact as long-time ``anomalous'' time-scaling of enstrophy leading to effective enhancement in the mixing capacity of the system.
 \end{itemize}

The structure of the rest of this work is outlined as follows: In section \ref{sec: Equtions}, we formulate the stochastic version of the Navier-Stokes equations for incompressible flows and develop our stochastic modeling procedure. In section \ref{sec: Comp_FrameWork}, we elaborate on the numerical methods we employ in our deterministic solver and generation of a proper grid and later on we introduce the our stochastic discretization approach followed by a discussion on how we study the significance of each source of stochasticty in a global sense. In section \ref{sec: Results}, we show the stochastic convergence, quantification of uncertainty in kinetic energy as QoI and we perform the global sensitivity analysis. Using the expected velocity and vorticity fields we computed from our stochastic computational framework, we obtain the fluctuating responses for a deterministic simulation and study their statistics in a qualitative and quantitative sense. Furthermore, we compute the enstrophy record associated with the fluctuating field and study its time-scaling that unravels a tied link between the observed highly non-Gaussian features and memory effects induced by long-lived coherent vortex structures. Finally, in section \ref{sec: Conclusion}, we point out the remarks of the present work and conclude our investigations.

\section{Stochastic Navier-Stokes Equations}\label{sec: Equtions}

Let $\Omega \subset \mathbb{R}^2$ be our bounded convex 2-D spatial domain with boundaries $\partial \Omega$.
Moreover, let $(\Omega_s, \mathcal{F},\mathbb{P})$ be a complete probability
space, where $\Omega_s$ is the space of events, $\mathcal{F} \subset 2^{\Omega_s}$ denotes the $\sigma$-algebra of sets in $\Omega_s$, and $\mathbb{P}$ is the probability measure. Then, the governing
stochastic incompressible 2-D Navier-Stokes (NS) equations subject to the continuity equation, $\nabla \cdot \pmb{V} = 0$, for Newtonian viscous fluids
\begin{align}\label{eqn: NS}
 \frac{\partial \pmb{V}}{\partial t} + \pmb{V} \cdot \nabla \pmb{V} = -\nabla p + \nu \nabla^2 \pmb{V}&, \qquad \forall (\mathbf{x}, t; \omega) \in \Omega \times (0,T] \times \Omega_s, \\ \nonumber
     \pmb{V}(\mathbf{x},t;\omega)=\pmb{V}_{\partial\Omega}&, \qquad \forall (\mathbf{x},t;\omega) \in \partial\Omega \times (0,T] \times \Omega_s, \\ \nonumber
     \pmb{V}(\mathbf{x},0;\omega)=\pmb{V}_0&, \qquad \forall (\mathbf{x};\omega) \in \Omega \times \Omega_s,
\end{align}
hold $\mathbb{P}$-almost surely subject to the corresponding proper initial and boundary conditions, introduced and modeled below. Here, $\pmb{V}(\mathbf{x}, t; \omega)$ represents vector of the velocity field for the fluid, $p(\mathbf{x}, t; \omega)$ denotes the specific pressure (including the density), and $\nu$ is the kinematic viscosity.

\subsection{Stochastic Modeling}\label{subsec: Stochastic_Modeling}

We are interested in learning how the symmetry-breaking factors would affect the onset of flow instability. In our modeling, these factors are reflected in terms of stochastic initial and boundary conditions, subsequently, the rest of possible random effects are treated deterministically. Accordingly, these symmetry-breaking effects are modeled through imposing a time-dependent wall angular velocity,
\begin{align}\label{eqn: Ang-vel}
    \dot{\theta}(t;\omega)&=\cos{\left(\alpha(\omega)  t\right)} \ e^{-\lambda(\omega) t}, \qquad \forall (t;\omega) \in (0,T] \times \Omega_s,
\end{align}
while we consider an off-centered rotation with a radial eccentricity of $\epsilon(\omega)$, $\forall \omega \in \Omega_s$, with respect to the geometric centroid of the cylinder. In our model, $\alpha(\omega)$ and $\lambda(\omega)$ denote the frequency of oscillations and the decay rate appearing in the angular velocity model, respectively. In other words, the no-slip boundary condition at the wall is imposed by the proposed wall velocity for which the initial condition is a solid-body and off-centered rotation. In our non-dimensional mathematical setup, the initial angular velocity, $\dot{\theta}(0;\omega)$, and the radius of the cylinder, $R$, are both taken to be unity. Therefore, the stochastic wall velocity field is expressed as
\begin{align}\label{eqn: BC}
    \pmb{V}_{\partial\Omega}(\mathbf{x},t;\omega) = \left( \mathbf{x} - \mathbf{r}_{\epsilon(\omega)} \right) \dot{\theta}(t;\omega), & \qquad \forall (\mathbf{x},t;\omega) \in \partial\Omega \times (0,T] \times \Omega_s, \\ \nonumber
    \Vert \mathbf{x} \Vert_2 = 1, & \quad \Vert \mathbf{r}_{\epsilon(\omega)}\Vert_2 = \epsilon(\omega),
\end{align}
where $\Vert \cdot \Vert_2$ denotes the $L^2$ norm.

\subsection{Parametrization of Random Space}\label{subsec: Parameterization}

Let $Y:\Omega_s \rightarrow \mathbb{R}^3$ be the set of independent random parameters, given as
\begin{align}\label{eqn: rnd_para}
    Y(\omega) = \lbrace Y_i \rbrace_{i=1}^3 = \lbrace \lambda(\omega), \alpha(\omega), \epsilon(\omega) \rbrace, \qquad \forall \omega \in \Omega_s,
\end{align}
with probability density functions (PDF) of each random parameter being $\rho_i: \Psi_i \rightarrow \mathbb{R}$, $i=1,2,3$, where $\Psi_i \equiv Y_i(\Omega_s)$ represent their images that are bounded intervals in $\mathbb{R}$. By independence, the joint PDF, $\rho(\pmb{\xi}) = \prod_{i=1}^3 \rho_i(Y_i), \quad \forall \pmb{\xi} \in \Psi$, with the support $\Psi = \prod_{i=1}^3 \Psi_i \subset \mathbb{R}^3$ form a mapping of the random sample space $\Omega_s$ onto the target space $\Psi$. Thus, an arbitrary point in the parametric space is denoted by $\pmb{\xi} = \lbrace \xi^1, \xi^2, \xi^3 \rbrace \in \Psi$. According to the Doob-Dynkin lemma \cite{oksendal2013stochastic}, we are allowed to represent the velocity field $\pmb{V}(\mathbf{x},t;\omega)$ as $\pmb{V}(\mathbf{x},t;\pmb{\xi})$, therefore, instead of working with the abstract sample space, we rather work in the target space. Finally, the formulation of stochastic governing equations in \eqref{eqn: NS} subject to the boundary/initial conditions in equation \eqref{eqn: BC} can be posed as: Find $\pmb{V}(\mathbf{x},t;\pmb{\xi}): \Omega \times (0,T] \times \Psi \rightarrow \mathbb{R}$ such that
\begin{align}\label{eqn: NS_Stch}
    \frac{\partial \pmb{V}}{\partial t} + \pmb{V} \cdot \nabla \pmb{V} &= -\nabla p + \nu \nabla^2 \pmb{V}, \\ \nonumber
    \pmb{V}(\mathbf{x},t;\pmb{\xi})=\pmb{V}_{\partial\Omega}, & \qquad \forall (\mathbf{x},t;\pmb{\xi}) \in \partial\Omega \times (0,T] \times \Psi, \\ \nonumber
    \pmb{V}(\mathbf{x},0;\pmb{\xi})=\pmb{V}_0, & \qquad \forall (\mathbf{x};\pmb{\xi}) \in \Omega \times \Psi,
\end{align}
hold $\rho$-almost surely for $\pmb{\xi}(\omega)\in \Psi$ and $\forall(\mathbf{x},t) \in  \Omega \times (0,T]$ subject to the incompressibility condition, $\nabla \cdot \pmb{V} = 0$.

\section{Stochastic Computational Fluid Dynamics Framework}\label{sec: Comp_FrameWork}

\subsection{Discretization of Physical Domain and Time-Integration}\label{subsec: Numerics}

Spectral/$hp$ element method \cite{karniadakis2013spectral} is a high-order numerical method to discretize the governing equations \eqref{eqn: NS} in the deterministic physical domain $\Omega$. In particular, SEM is a proper candidate to achieve a high-order accuracy discretization close to the wall boundaries. In SEM, we partition the spatial domain, $\Omega$, into non-overlapping elements as $\Omega = \bigcup_{e=1}^{N_{el}} \Omega^e$, where $N_{el}$ denotes the total number of elements in $\Omega$. In practice, a standard element, $\Omega^{st}$, is constructed in a way that its local coordinate, $\pmb{\zeta} \in \Omega^{st}$, is mapped to the global coordinate for any elemental domain, $\mathbf{x} \in \Omega^e$. This mapping is performed through an iso-parametric transformation, $\mathbf{x} = \chi^e(\pmb{\zeta})$. Within the standard element, a polynomial expansion of order $P$ is employed to represent the approximate solution, $V^\delta$, as
\begin{align}\label{eqn: SEM}
    V^\delta(\mathbf{x}) = \sum_{e=1}^{N_{el}}\sum_{j=1}^P \hat{V}_j^e \Phi_j^e(\pmb{\zeta}) = \sum_{i=1}^{N_{dof}} \hat{V}_i \Phi_i(\mathbf{x}),
\end{align}
where $N_{dof}$ indicates the total degrees of freedom (DoF) \textit{i.e.}, the modal coefficients in the solution expansion. Moreover, $\Phi_j^e(\pmb{\zeta})$ are the local expansion modes, while $\Phi_i(\mathbf{x})$ are the global modes that are obtained from the global assembly procedure of the local modes \cite{karniadakis2013spectral}.

\texttt{NEKTAR++} \cite{CANTWELL2015, moxey2020nektar++}, a parallel open-source numerical framework, provides a seamless platform offering efficient implementation of multiple SEM-based solvers in addition to the pre-/post-processing tools. In our study, we employ its incompressible Navier-Stokes solver namely as \texttt{IncNavierStokesSolver}. Here, the velocity correction scheme along with the $C^0$-continuous Galerkin projection are utilized as splitting/projection method in order to decouple the velocity and the pressure fields \cite{CANTWELL2015}. We use $P$th-order polynomial expansions \textit{i.e.}, the modified Legendre basis functions while we vary $P$ for elements at different spatial regions (see section \ref{subsubsec: Grid}). Moreover, a second-order implicit-explicit (IMEX) time-integration scheme is used while the time-step is fixed during the time-stepping. The spectral vanishing viscosity (SVV) technique \cite{KIRBY2006_SVV, karniadakis2013spectral} is also used to ensure a stabilized numerical solution from spectral/$hp$ element method.

\subsubsection {Grid Generation}\label{subsubsec: Grid}

\begin{figure}[t!]
    \begin{minipage}[b]{.49\linewidth}
        \centering
        \includegraphics[width=\textwidth]{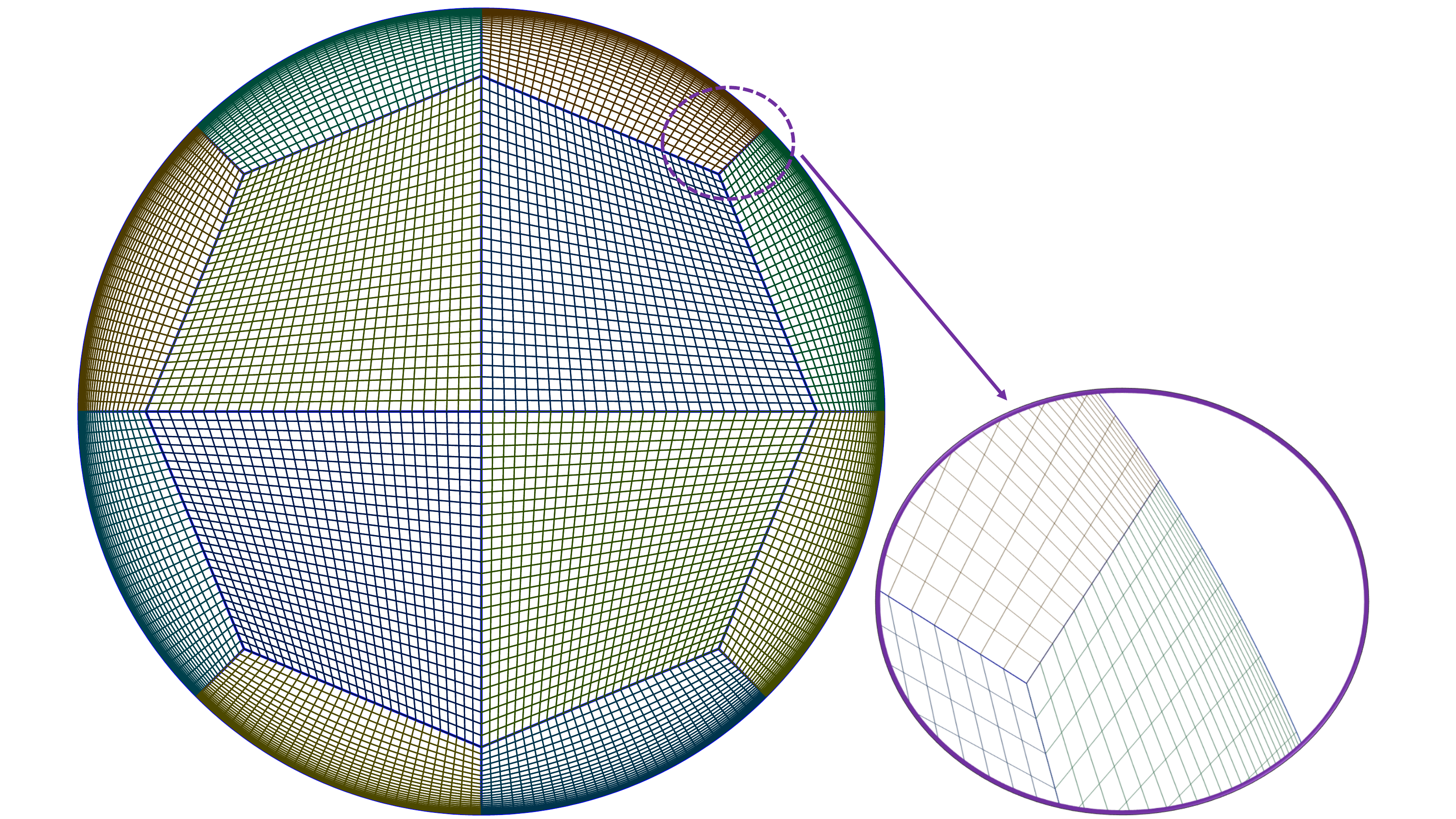}
        \subcaption{Generated structured grid with transitional $h$-refinement.}\label{fig: hGrid}
    \end{minipage}
    \begin{minipage}[b]{.49\linewidth}
        \centering
        \includegraphics[width=\textwidth]{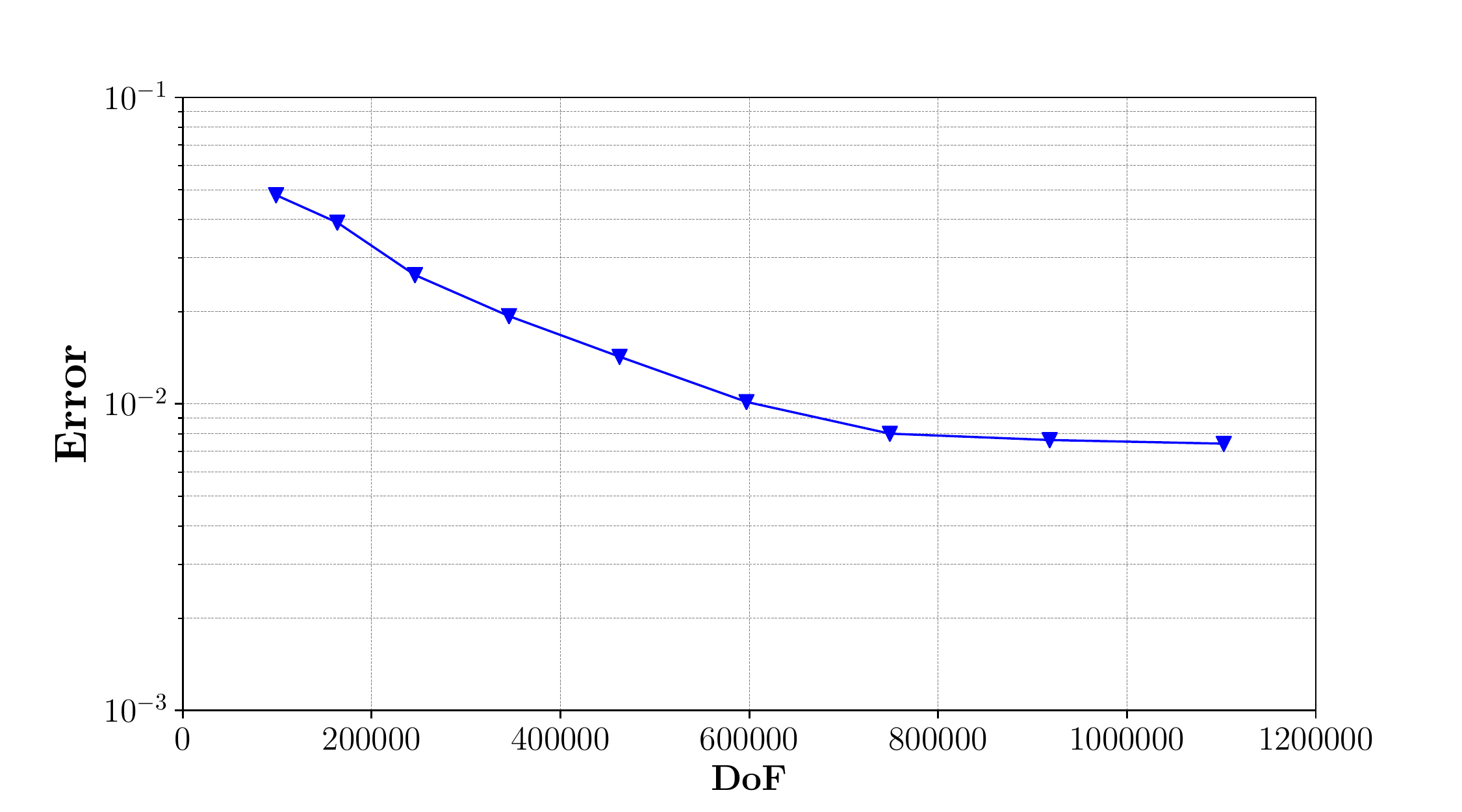}
        \subcaption{Grid convergence study based on the error in kinetic energy.}\label{fig: GridConv}
    \end{minipage}
    \caption{Constructed grid and the analysis of grid-independent solution.}\label{fig: Grid}
\end{figure}

A 2-D structured grid is generated with quadrilateral elements considering $h$-type refinement technique to attain proper grid resolution near the wall. We utilize the open-source finite element grid generator, \texttt{Gmsh} \cite{geuzaine2009gmsh}, to construct the geometry and then the $h$-refined grid. The generated grid is illustrated in Figure \ref{fig: hGrid}, which shows elemental nodes and $h$-refinement near the wall. For this $h$-refined grid, we employ a spatially-variable polynomial expansion \cite{moxey2020nektar++} so that we gain high-accuracy close to wall, while avoiding unnecessary computational cost away from the wall. In order to ensure that our solution is independent of the grid resolution for the Reynolds number, $Re = R^2 \dot{\theta}/\nu = \dot{\theta}/\nu$, fixed at $10^6$, we carry out a grid convergence study based on the error we obtain from the difference of the time-integrated kinetic energy between the solutions after varying the grid resolution and a reference solution with $\sim 2.1\times 10^6$ total DoF. As shown in Figure \ref{fig: GridConv}, the total DoF of $\sim 7.5 \times 10^5$ gives us a sufficient grid resolution ensuring that the numerical solution is independent of grid resolution. In the applied IMEX time-integration scheme, the time-step is fixed at $\Delta t=4\times10^{-5}$ while the numerical stability is always checked during the simulations by ensuring that CFL number being less than unity. In particular, our SEM grid is achieved by utilizing 9th-order polynomial expansions for the elements in the near the wall region and 7th-order polynomial expansions for the elements in the cylinder's core region. In other words, due to this spatial $p$-refinement procedure, the near-wall elements would consist of 64 rectangular sub-elements ($P=9$) and, the elements in the core region will be finer 36 times ($P=7$). For flow at moderately low Reynolds numbers, we verify the resulting solutions from our numerical setup through a comparison with analytical solutions (see Appendix \ref{sec: Appendix1}).

\subsection{Stochastic Discretization}\label{subsec: Stoch_FrameWork}

Sampling from the parametric random space introduced in section \ref{subsec: Parameterization} is a non-intrusive approach for stochastic discretization. Monte Carlo (MC) sampling method is the most conventional way to perform such task, however, the large number of required realizations of random space is its bottleneck, which prohibits utilizing MC for computationally demanding problems. In our study, we employ probabilistic collocation method (PCM) \cite{xiu2005high, nobile2008sparse, foo2008multi}, which is a non-intrusive scheme and has shown affordable efficiency by providing fairly fast convergence rate for statistical moments. In PCM, a set of collocation points $\lbrace \pmb{q}_j \rbrace_{j=1}^{\mathcal{J}}$ is prescribed in parametric random space $\Psi$, where $\mathcal{J}$ denotes the number of collocation points. As a common practice to construct a stable basis, $\lbrace \pmb{q}_j \rbrace_{j=1}^{\mathcal{J}}$ are taken to be the points of a suitable cubature rule on $\Psi$ with integration weights, $\lbrace \pmb{w}_j \rbrace_{j=1}^{\mathcal{J}}$. In this work, we employ a fast algorithm proposed by Glaser \textit{et al.} \cite{Glaser2007_GLR} to compute the collocation points based on Gauss quadrature rule. Therefore, let the solution $\pmb{V}$ in the parametric random space be collocated on the set of points $\lbrace \pmb{q}_j \rbrace_{j=1}^{\mathcal{J}}$. In other words, we use the SEM setup described in section \ref{subsec: Numerics} to solve a set of deterministic problems in which the wall velocity field $\pmb{V}_{\partial \Omega}(\mathbf{x},t; \pmb{\xi})$ in equation \eqref{eqn: NS_Stch} is replaced with its deterministic realization $\pmb{V}_{\partial \Omega}(\mathbf{x},t; \pmb{q}_j)$. In order to construct the approximate stochastic solution $\hat{\pmb{V}}(\mathbf{x},t; \pmb{\xi})$ from a set of deterministic solutions $\lbrace \pmb{V}(\mathbf{x},t; \pmb{q}_j)\rbrace_{j=1}^{\mathcal{J}}$, we employ $L_i(\pmb{\xi})$, the Lagrange interpolation polynomials of order $i$. Let $\mathcal{I}$ represent the approximation operator, therefore, the approximate stochastic solution is written as
\begin{align}\label{eqn: uhat}
    \hat{\pmb{V}}(\mathbf{x},t; \pmb{\xi}) = \mathcal{I} \pmb{V}(\mathbf{x},t; \pmb{\xi}) = \sum_{j=1}^\mathcal{J} \pmb{V}(\mathbf{x},t; \pmb{q}_j)L_j(\pmb{\xi}).
\end{align}
We choose the approximation operator $\mathcal{I}$ to be the full tensor product of the Lagrange interpolants in each dimension of parametric random space.
Defining the PDF $\rho(\pmb{\xi})$ over the parametric random space and using the approximate solution, the expectation of $\pmb{V}$ is computed as
\begin{align}\label{eqn: int1}
    \mathbb{E}\left[\pmb{V}(\mathbf{x},t;\pmb{\xi})\right]=\int_{\Psi} \hat{\pmb{V}}(\mathbf{x},t; \pmb{\xi}) \rho(\pmb{\xi}) d\pmb{\xi}.
\end{align}
This integral would be approximated using a proper quadrature rule. Letting the set of interpolation/collocation points $\lbrace \pmb{q}_j \rbrace_{j=1}^{\mathcal{J}}$ obtained from Glaser \textit{et al.} \cite{Glaser2007_GLR} coincide these quadrature points with associated integration weights $\lbrace \pmb{w}_j \rbrace_{j=1}^{\mathcal{J}}$, one can efficiently compute the approximation to the integral in equation \eqref{eqn: int1}. Applying the Kronecker delta property of Lagrange interpolants, this integral is approximated as
\begin{align}\label{eqn: int2}
    \mathbb{E}\left[\pmb{V}(\mathbf{x},t;\pmb{\xi})\right] \approx \sum^{\mathcal{J}}_{j=1} w_j \, \rho(\pmb{q}_j) \; \pmb{J} \; \pmb{V}(\mathbf{x},t;\pmb{q}_j).
\end{align}
In equation \eqref{eqn: int2}, $\pmb{J}$ represents the Jacobian associated with an affine mapping from standard to the real integration domain regarding the applied quadrature rule. In our study, we utilize uniformly distributed random variables to represent symmetry-breaking effects, hence, $\pmb{J} \, \rho(\pmb{q}_j)$ yields a constant. In the case of our problem with three stochastic dimensions, $\pmb{J} \, \rho(\pmb{q}_j)=(\frac{1}{2})^3$. Consequently, the approximate computation of the expectation integral \eqref{eqn: int1} is simplified to
\begin{align}\label{eqn: EX}
    \mathbb{E}[\pmb{V}(\mathbf{x},t;\pmb{\xi})] \approx \frac{1}{8}\sum^{\mathcal{J}}_{j=1} w_j \pmb{V}(\mathbf{x},t;\pmb{q}_j).
\end{align}
Similar to the MC approach and using \eqref{eqn: EX}, the standard deviation in our problem is approximated as
\begin{align}\label{eqn: standard1}
    \pmb{\sigma}\left[\pmb{V}(\mathbf{x},t;\pmb{\xi})\right] \approx \sqrt{\frac{1}{8}\sum^{\mathcal{J}}_{j=1} w_j \Big(\pmb{V}(\mathbf{x},t;\pmb{q}_j)-\mathbb{E}[\pmb{V}(\mathbf{x},t;\pmb{\xi})]\Big)^2}.
\end{align}

\section{Variance-Based Sensitivity Analysis}\label{sec: Var-based_SA}

Grasping knowledge on the significance of sources of randomness in a stochastic modeling procedure could be very helpful in terms of reducing the computational cost and also decision making during stochastic modeling.Variance-based sensitivity analysis is a well-known technique to assess the relative effect of randomness in each stochastic dimension on the total variance of any QoI, $U$, as the output of a stochastic model in a global sense \cite{SOBOL2001,SALTELLI2010}. In practice, sensitivity of the QoI to each stochastic parameter is measured by the conditional variance in the QoI, which is caused by that specific parameter. In general, for a $k$-dimensional stochastic space, $\pmb{\xi}$, a QoI may be represented as a square-integrable function of the stochastic parameters $U=f\left(\pmb{\xi}\right)$. Using Hoeffding decomposition of $f$ \cite{SOBOL1990}, and also the conditional expectation of the stochastic model, $\mathbb{E}\left[U\vert\xi^i\right] \ _{(i=1,\dots,k)}$, the total variance of $U$ can be decomposed as
\begin{align}\label{eqn: Var_decompose}
    V(U) = \sum_i V_i + \sum_i\sum_{i<j}V_{ij} + \dots + V_{12\dots k},
\end{align}
where $V_i$ and $V_{ij}$ are represented by
\begin{align}\label{eqn: 1st_2nd_Var}
    V_i =& V_{\xi^i}\Big(\mathbb{E}_{\pmb{\xi}^{\sim i}}\left[U\vert \xi^i \right]\Big), \\ \nonumber
    V_{ij} =& V_{\xi^i\xi^j}\Big(\mathbb{E}_{\pmb{\xi}^{\sim ij}}\left[U\vert \xi^i,\xi^j\right]\Big) - V_{\xi^i}\Big(\mathbb{E}_{\pmb{\xi}^{\sim i}}\left[U\vert \xi^i \right]\Big) - V_{\xi^j}\Big(\mathbb{E}_{\pmb{\xi}^{\sim j}}\left[U\vert \xi^j \right]\Big).
\end{align}
Similarly, the higher order terms, $V_{i_1 i_2 \dots i_n}$, $_{n\leq k}$, are defined. In equation \eqref{eqn: 1st_2nd_Var}, $V_{\xi^i}\Big(\mathbb{E}_{\pmb{\xi}^{\sim i}}\left[U\vert \xi^i \right]\Big)$ is representing the first-order effects of $\xi^i$ on the total variance of QoI, $V(U)$, and $\pmb{\xi}^{\sim i}$ indicates the set of all the stochastic parameters excluding $\xi^i$ that is assumed to be fixed. Moreover, $V_{\xi^i\xi^j}\Big(\mathbb{E}_{\pmb{\xi}^{\sim ij}}\left[U\vert \xi^i,\xi^j\right]\Big)$ denotes the joint effects of stochasticity in $\xi^i$ and $\xi^j$ on the total variance. In general, $\mathbb{E}_{\pmb{\xi}^{\sim ij\dots}}\left[U\vert \xi^i, \xi^j, \dots \right]$ is the expectation of $U$, which is taken over all values of $\pmb{\xi}^{\sim ij\dots}$, while the stochastic parameters $(\xi^i, \xi^j, \dots)$ are fixed at specific values, hence, $V_{\xi^i\xi^j\dots}\Big(\mathbb{E}_{\pmb{\xi}^{\sim ij\dots}}\left[U\vert \xi^i,\xi^j,\dots \right]\Big)$ gives the reduction in total variance.

According to the law of total variance, one can decompose the total variance of $U$ by conditioning on one specific stochastic parameter such as $\xi^i$ as follows
\begin{align}\label{eqn: Total_Var_Si}
    V(U) = V_{\xi^i}\Big(\mathbb{E}_{\pmb{\xi}^{\sim i}}\left[U\vert \xi^i \right]\Big) + \mathbb{E}_{\xi^i}\Big(V_{\pmb{\xi}^{\sim i}}\left[U\vert \xi^i \right]\Big),
\end{align}
where $\mathbb{E}_{\xi^i}\Big(V_{\pmb{\xi}^{\sim i}}\left[U\vert \xi^i \right]\Big)$ represents the residual of the total variance. By normalizing the first term in the right-hand side of equation \eqref{eqn: Total_Var_Si}, we can obtain the global sensitivity indices, namely, \textit{Sobol} indices \cite{SOBOL2001}
\begin{align}\label{eqn: Sobol_indx}
    S^i=\frac{V_{\xi^i}\Big(\mathbb{E}_{\pmb{\xi}^{\sim i}}\left[U\vert \xi^i \right]\Big)}{V\left(U\right)},
\end{align}
where, $S^i$ determines the first-order contribution of $\xi^i$ in the random parameter space on the total variance of the QoI is considered, hence, no joint contributions embedded in the residual term is taken into account. 

\section{Numerical Results}\label{sec: Results}

\subsection{Stochastic Convergence and Uncertainty Quantification}\label{subsec: Stoch_conv_UQ}

We seek to attain the number of required number of collocation points (PCM realizations) in order to have a converged solution for the first-order and second-order moments, \textit{i.e.}, expectation and variance, respectively. This is a crucial step to ensure that the propagated parametric uncertainty that is embedded in the stochastic model (described in section \ref{sec: Equtions}) is properly captured and quantified regardless of the total number of realizations (forward solutions) we use in PCM. The aforementioned parametric uncertainty, as defined in section \ref{sec: Equtions}, $\pmb{\xi}=\lbrace \xi^1,\xi^2,\xi^3 \rbrace$, and the distributions associated with each parameter is reported in Table \ref{Table: stochastic_parameters}.

\begin{figure}[t!]
 \centering
  \includegraphics[width=.8\textwidth]{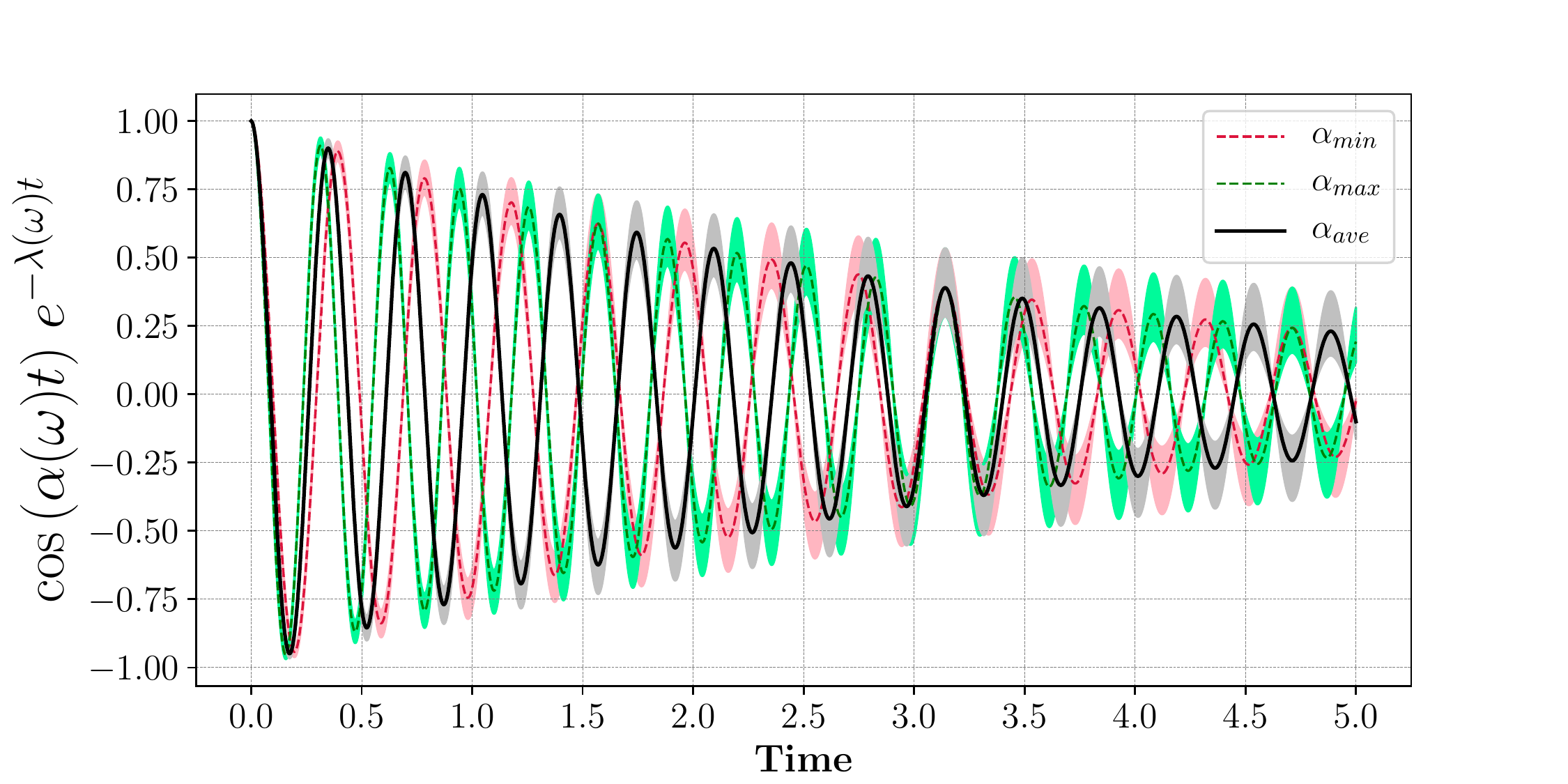}
  \caption{Stochastic angular velocity, $\dot{\theta}(t;\omega)$, including the decay, $\lambda(\omega)$, and oscillatory, $\alpha(\omega)$, effects with respect to Table \ref{Table: stochastic_parameters}. The colored bounds illustrate the variability of angular velocity for the depicted realizations of $\alpha$.}\label{fig: Ang_Vel}
\end{figure}

\begin{table}[t!]
	\centering
	\caption{Stochastic parameters of the wall velocity model and their mean values.}
	\label{Table: stochastic_parameters} 
	\begin{tabular}{l l l}
		\hline \hline
		Stochastic parameter & $\qquad$ & Distribution \\
		\hline
		$\xi^1:$ (decay rate) & $\qquad$ & $\sim \mathcal{U}\left(0.2,0.4\right)$ \\
		$\xi^2:$ (oscillations' frequency) & $\qquad$ & $\sim \mathcal{U}\left(16,20\right)$ \\
		$\xi^3:$ (eccentricity of rotation) & $\qquad$ & $\sim \mathcal{U}\left(0,0.05\right)$ \\
		\hline \hline
	\end{tabular}
	\vspace{0.1 in}
\end{table}
According to Table \ref{Table: stochastic_parameters}, the resulting randomness in the angular velocity is shown in Figure \ref{fig: Ang_Vel}.
 For a three-dimensional random space regarding our stochastic model and considering a full tensor product PCM we want to evaluate the stochastic behavior and also uncertainty propagation in the dynamics of flow. By choosing the kinetic energy, $E(t)$, as QoI, we perform the stochastic convergence study while we keep increasing the number of collocation points in all stochastic directions. It is worth mentioning that kinetic energy is a fair candidate as QoI since it represents the dynamics of the entire system without being biased towards a specific spatial direction or location. The kinetic energy is defined as:
\begin{align}\label{eqn: KE}
    E(t) = \frac{1}{2\mu(\pmb{\Omega})}\int_{\pmb{\Omega}} \Vert \pmb{V} \Vert^2 d\pmb{\Omega},
\end{align}
where $\mu(\pmb{\Omega})$ denotes the area of the spatial domain, $\pmb{\Omega}$, and $\Vert \pmb{V} \Vert$ represents the $L^2$ norm of velocity field. 

\begin{figure}[t!]
    \centering
    \includegraphics[width=.8\textwidth]{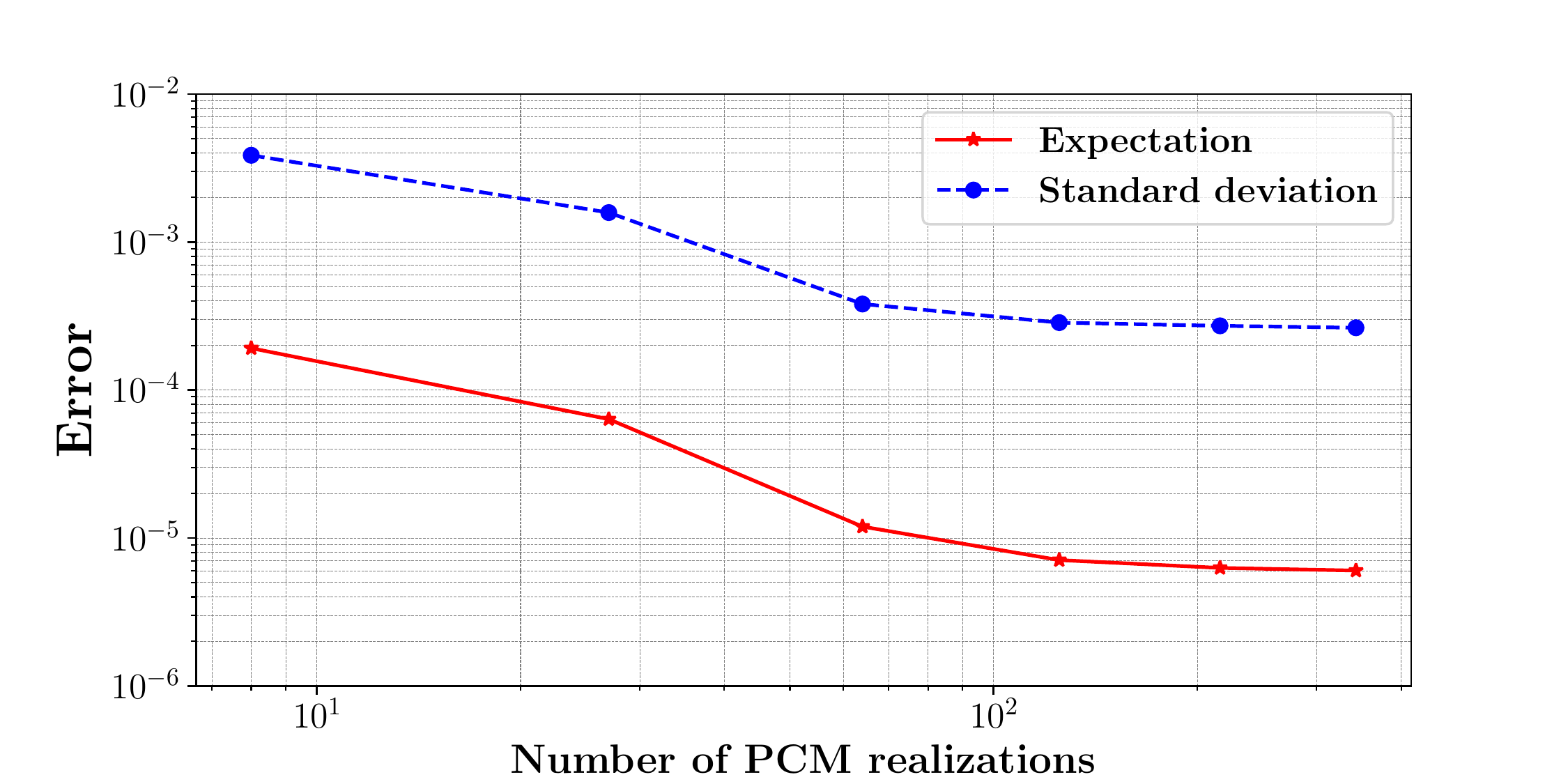}
    \caption{Stochastic convergence study for PCM considering expectation and standard deviation of the kinetic energy. The reference solution to compute the errors comes from a expectation and standard deviation of kinetic energy computed from a 2500 MC samples of random space.}\label{fig: stochastic_conv}
\end{figure}

After post-processing the outputs of each realization, we have an array of kinetic energy, which is computed for the entire simulation time. The reference solution for the stochastic convergence study is the expectation and variance of kinetic energy obtained from a Monte Carlo approach with 2500 realizations that are initially generated from Latin Hypercube Sampling (LHS) of random space reported in Table \ref{Table: stochastic_parameters}. Thus, one can compute the error for expectation and standard deviation of kinetic energy while changing the number of PCM realizations by increasing the number of collocation points. As shown in Figure \ref{fig: stochastic_conv}, by taking five collocation points (125 PCM realizations) the expectation and standard deviation become independent of the number of collocation points, hence, the stochastic convergence is achieved.

\begin{figure}[t!]
    \begin{minipage}[b]{.32\linewidth}
        \centering
        \includegraphics[width=1\textwidth]{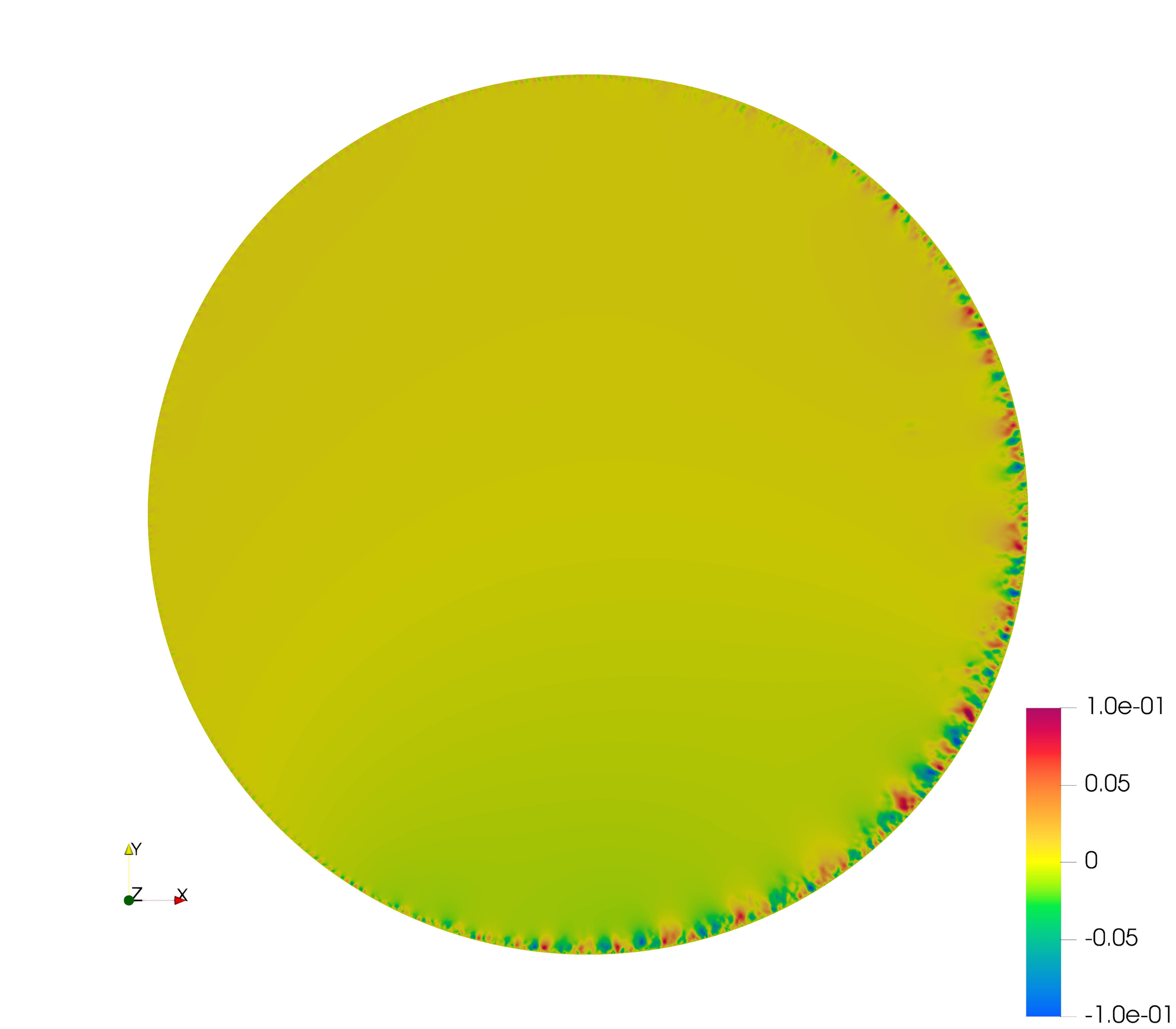}
        \subcaption{$\mathbb{E}\left[u_r(\mathbf{x};\pmb{\xi})\right]$, $t=2.5$}
    \end{minipage}
    \begin{minipage}[b]{.32\linewidth}
        \centering
        \includegraphics[width=1\textwidth]{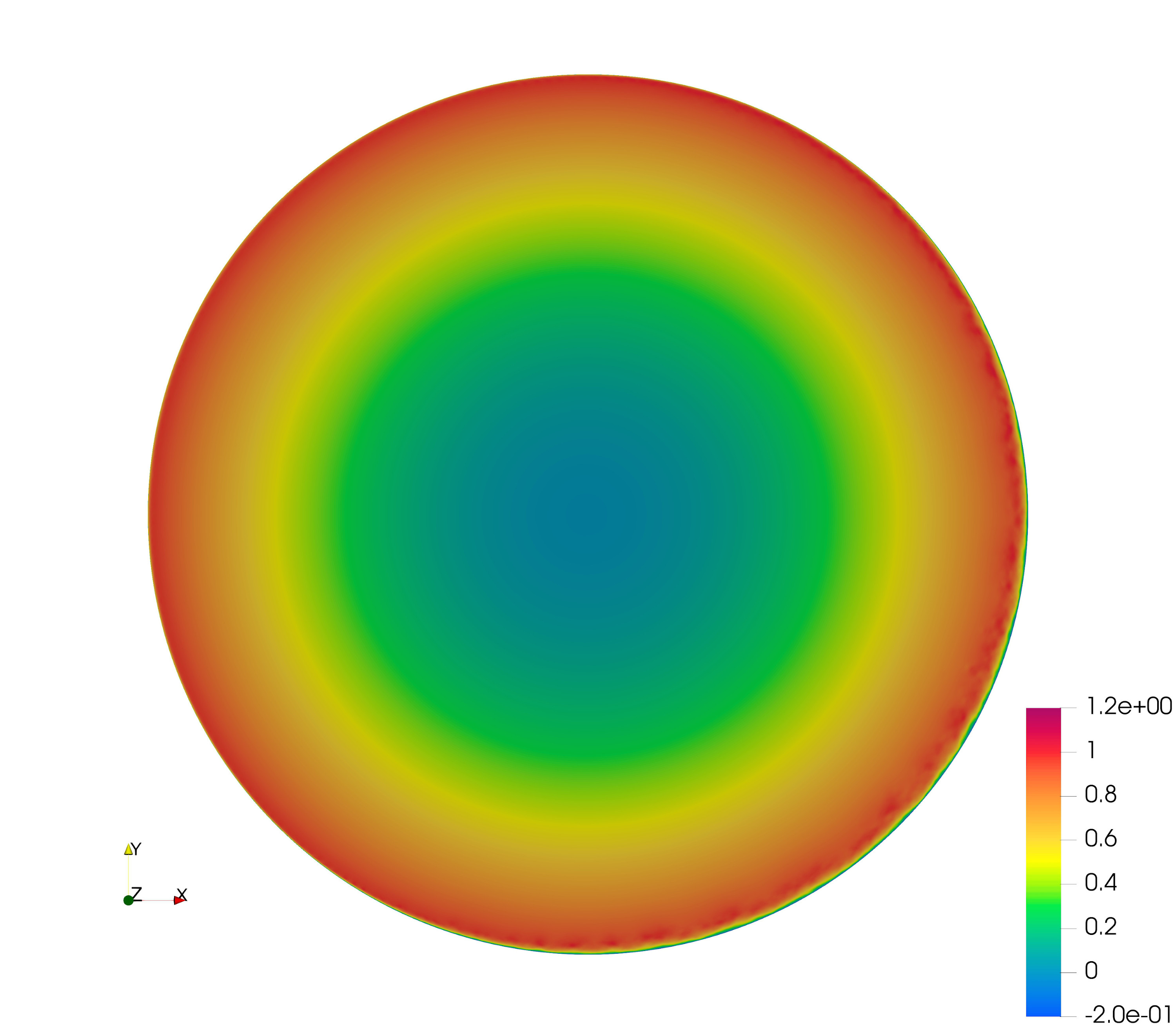}
        \subcaption{$\mathbb{E}\left[u_{\theta}(\mathbf{x};\pmb{\xi})\right]$, $t=2.5$}
    \end{minipage}
    \begin{minipage}[b]{.32\linewidth}
        \centering
        \includegraphics[width=1\textwidth]{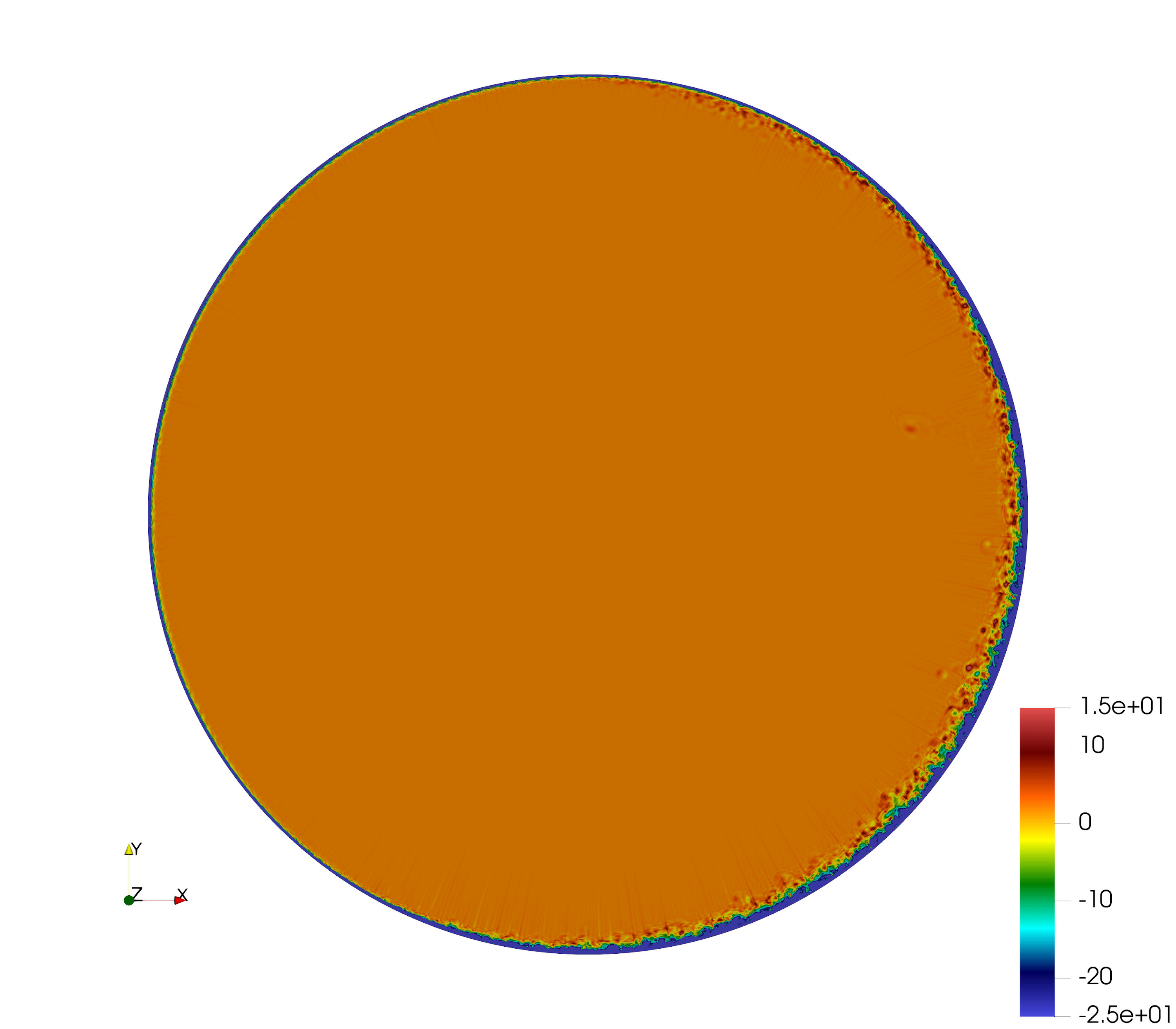}
        \subcaption{$\mathbb{E}\left[\omega_z(\mathbf{x};\pmb{\xi})\right]$, $t=2.5$}
    \end{minipage}    
    \begin{minipage}[b]{.32\linewidth}
        \centering
        \includegraphics[width=1\textwidth]{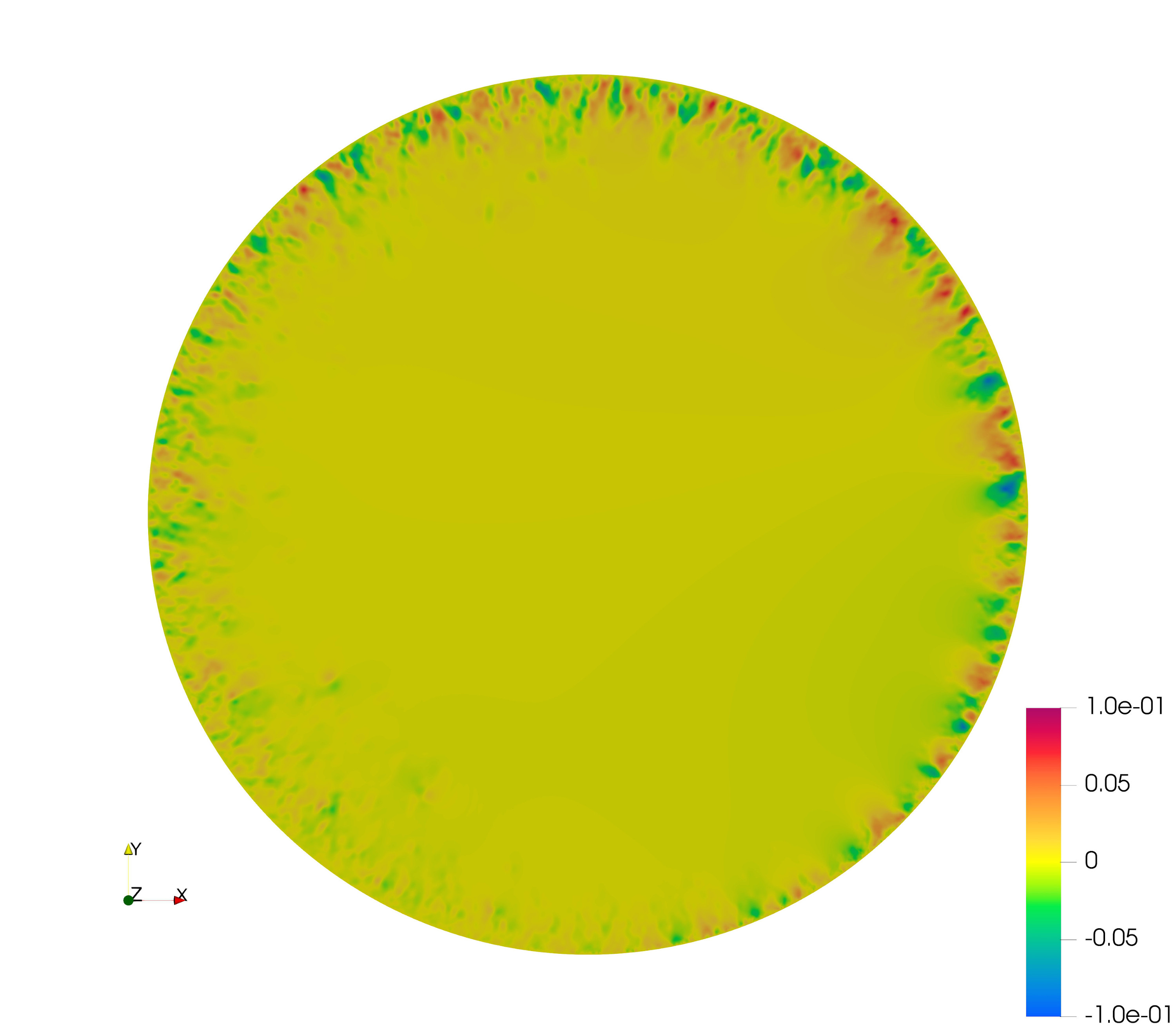}
        \subcaption{$\mathbb{E}\left[u_r(\mathbf{x};\pmb{\xi})\right]$, $t=5$}
    \end{minipage}
    \begin{minipage}[b]{.32\linewidth}
        \centering
        \includegraphics[width=1\textwidth]{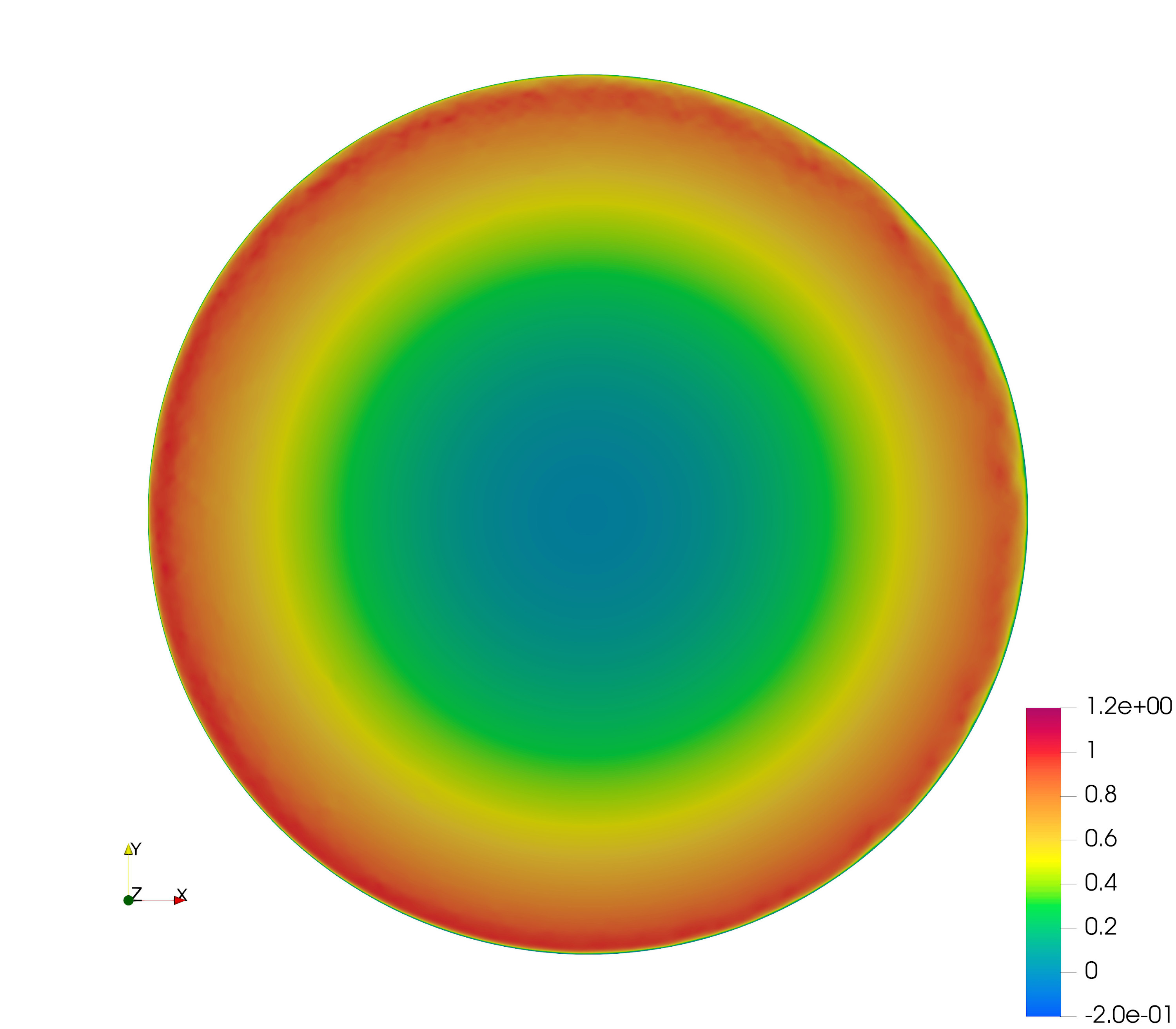}
        \subcaption{$\mathbb{E}\left[u_{\theta}(\mathbf{x};\pmb{\xi})\right]$, $t=5$}
    \end{minipage}
    \begin{minipage}[b]{.32\linewidth}
        \centering
        \includegraphics[width=1\textwidth]{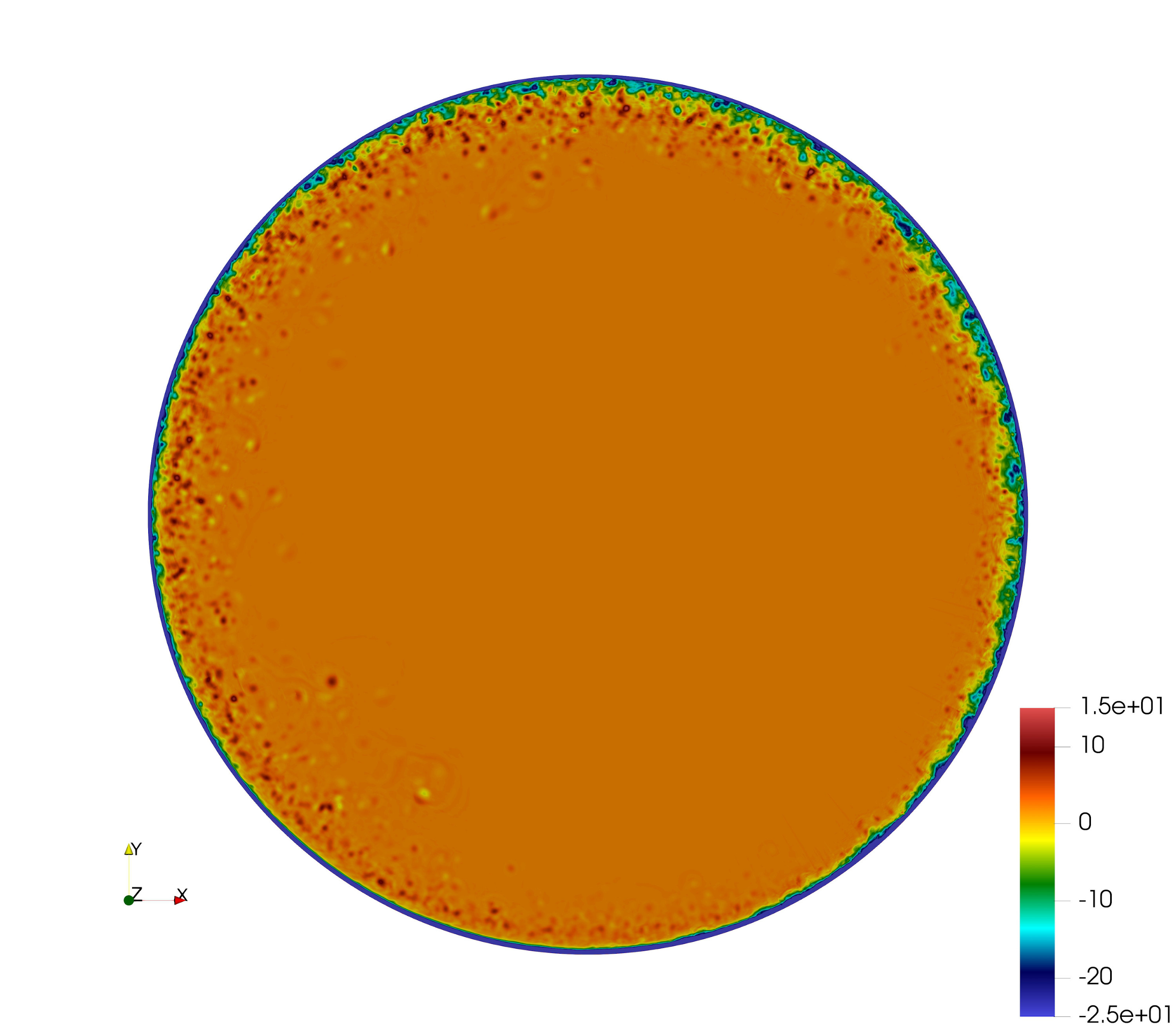}
        \subcaption{$\mathbb{E}\left[\omega_z(\mathbf{x};\pmb{\xi})\right]$, $t=5$}
    \end{minipage}
    \caption{Snapshots of expected velocity and vorticity fields obtained from converged PCM with 125 realizations.}\label{fig: vorticity_exp}
\end{figure}

Since the geometry of this flow is well-represented in the polar coordinate system $(r-\theta)$, we manage to transform the velocity field for the converged PCM case as $\pmb{V}=(u_r,u_\theta)$, which are derived as
\begin{align}\label{eqn: polar_vel}
    u_r = \frac{x u_x +y u_y}{r}, \qquad u_\theta = \frac{x u_y-y u_x}{r}.
\end{align}
where $u_x$ and $u_y$ represent velocity components along $x$ and $y$ directions in the Cartesian coordinate system, $r=\sqrt{x^2+y^2}$ is the radial location from cylinder center and $\theta$ denotes the azimuth angle. Having the velocity components transformed as equation \eqref{eqn: polar_vel}, Figure \ref{fig: vorticity_exp} portrays the snapshots of expected velocity components and also vorticity, $\omega_z=\partial u_y/\partial x - \partial u_x/\partial y$, fields at $t = 2.5$ and 5. 

\begin{figure}[t!]
    \begin{minipage}[b]{.49\linewidth}
        \centering
        \includegraphics[width=1\textwidth]{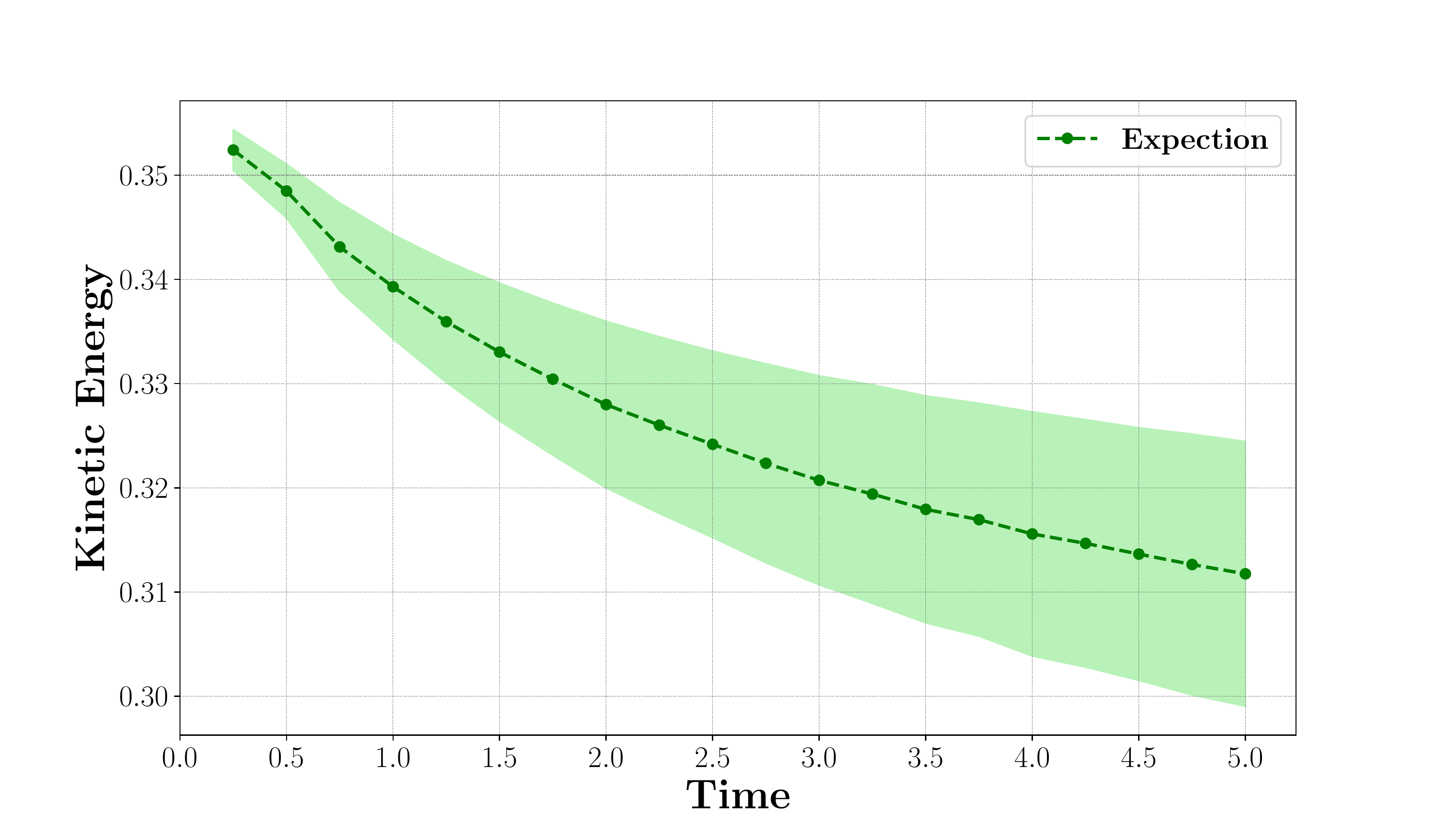}
        \subcaption{Expectation and uncertainty propagation}\label{fig: KE_evolution_UQ_a}
    \end{minipage}
    \begin{minipage}[b]{.49\linewidth}
        \centering
        \includegraphics[width=1\textwidth]{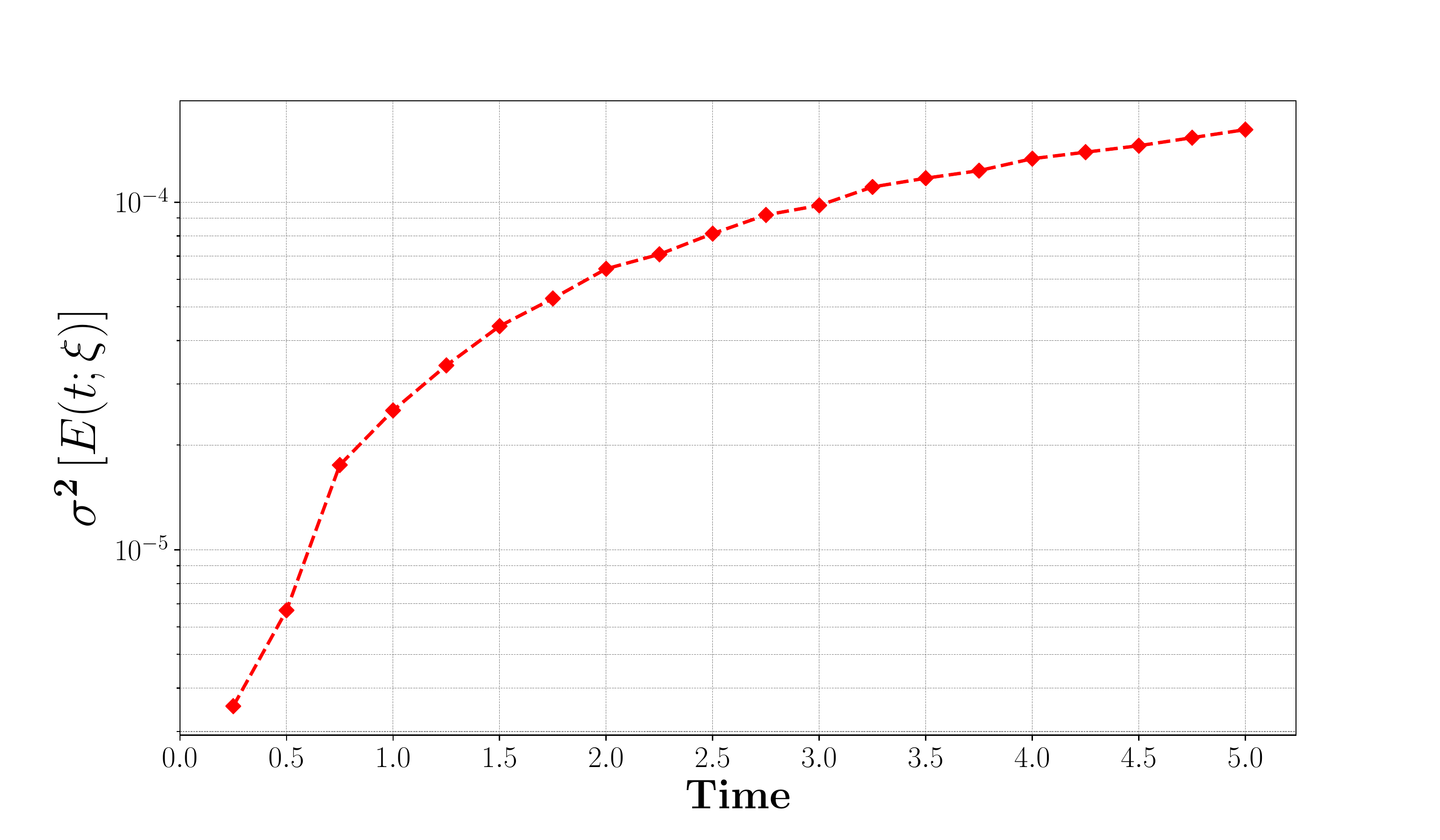}
        \subcaption{Time evolution of variance}\label{fig: KE_evolution_UQ_b}
    \end{minipage}
    \caption{Time evolution of expected kinetic energy and its uncertainty propagation where green colored area identifies the $\mathbb{E}\left[E(t;\pmb{\xi})\right]\pm \pmb{\sigma}\left[E(t;\pmb{\xi})\right]$. Here, $\pmb{\xi}$ represents the vector of parametric uncertainty in the random space, which is discretized with 125 PCM realizations.}\label{fig: KE_evolution_UQ}
\end{figure}

The regularity of the solution to the stochastic Navier-Stokes equations in the parametric space is a crucial point in the effective use of PCM \cite{maitre2010spectral}. Here, we assume that the solution is smooth enough of finite variance. Therefore, using the sufficiently converged PCM, which properly incorporates the effects of parametric uncertainty in our model, $\pmb{\xi}$, we can compute the time evolution of the expected value of kinetic energy, $\mathbb{E}\left[E(t;\pmb{\xi})\right]$. Moreover, it enables us to quantify the uncertainty, which is propagated with time through the kinetic energy as our dynamics-representative QoI \cite{ZAYERNOURI2013_roughness, KHALIL2015, Patra2018comparative, Kharazmi2020, DeMoraes2020, yangmethod}. Subsequently, Figure \ref{fig: KE_evolution_UQ} shows the time evolution of expected kinetic energy and the uncertainty bounds computed from $\mathbb{E}\left[E(t;\pmb{\xi})\right]\pm \pmb{\sigma}\left[E(t;\pmb{\xi})\right]$. Clearly, the propagation of uncertainty grows with time as we compare the uncertainty bounds at the onset of the instability with the later times, which is shown in Figure \ref{fig: KE_evolution_UQ_a}. Additionally, the rate of the uncertainty propagation might be learned by looking at the time evolution of kinetic energy variance $\pmb{\sigma}^2\left[E(t;\pmb{\xi})\right]$. Accordingly, Figure \ref{fig: KE_evolution_UQ_b} illustrates that the variance grows almost exponentially when $t<0.75$ and after a short transition time it grows linearly, therefore, the rate of the uncertainty propagation is much faster and more influential close to the onset of the instability.

\subsection{Sensitivity Analysis on Kinetic Energy}\label{subsec: SA_KE}

The focus of this section is to evaluate the effects of each stochastic parameter on the underlying variations of kinetic energy as the quantity of interest. The global sensitivity indices introduced in section \ref{sec: Var-based_SA} are proper measures to study the importance of each source of randomness on the dynamics of the symmetry-breaking flow instability, which was stochastically computed using PCM in previous section. Variance-based sensitivity analysis is usually performed by employing realizations of random space through Monte Carlo approach \cite{HAMDIA2015, VUBAC2015, FARRELL2015, Faghihi2018, DeMoraes2020}. However, here we are interested in using the solution of our stochastic convergence study (125 PCM realizations cases) to compute the expected variance reductions conditioned on $\xi^i$ according to equation \eqref{eqn: Sobol_indx} and, hence, the sensitivity indices, $S^i$. 

\begin{figure}[t!]
    \centering
    \includegraphics[width=.8\textwidth]{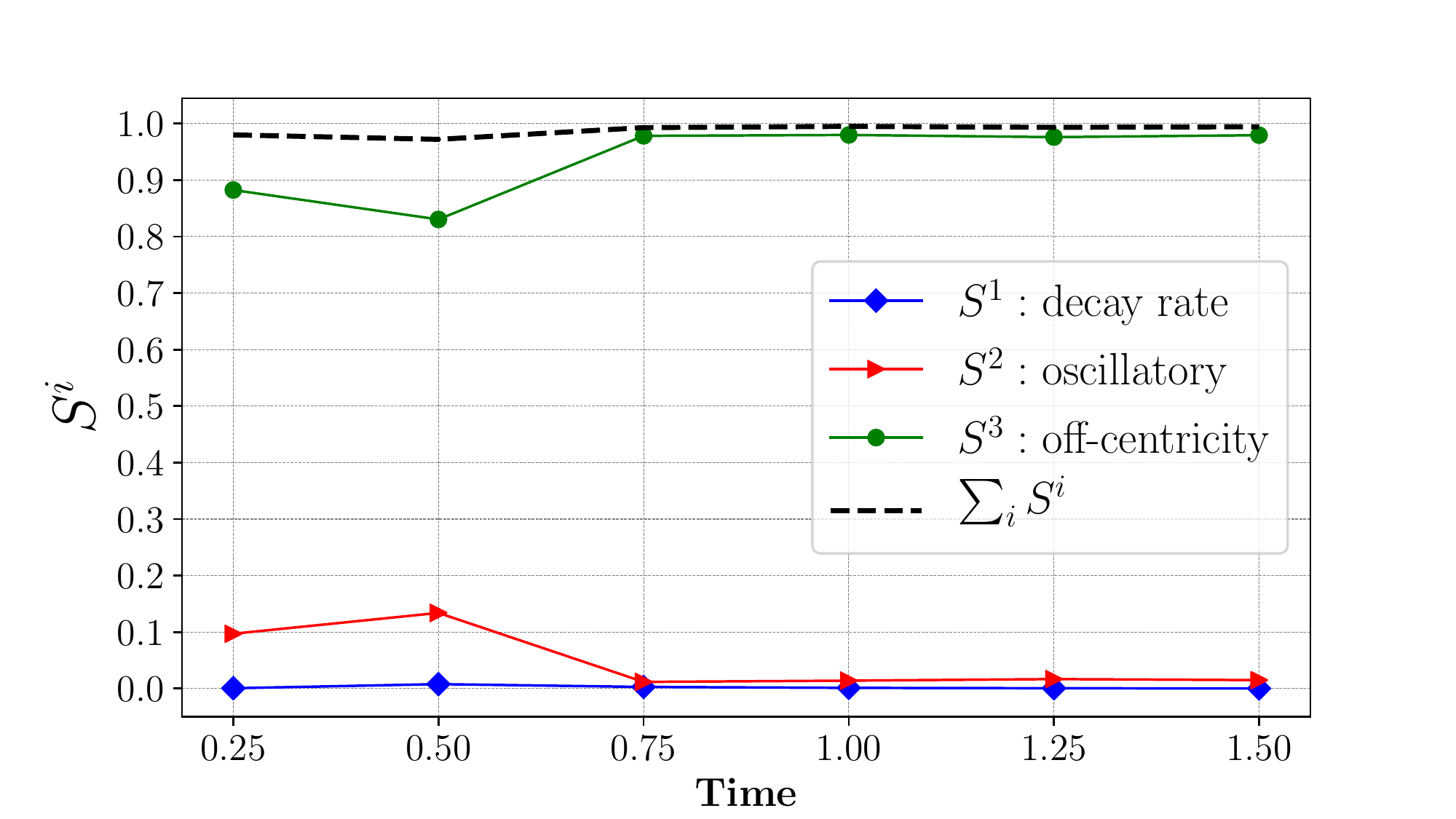}
    \caption{Time evolution of global sensitivity indices, $S^i$, for the stochastic parameters, $\pmb{\xi}$,  considering kinetic energy, $E(t;\pmb{\xi})$, as the QoI.}\label{fig: S_indc}
\end{figure}

Figure \ref{fig: S_indc} shows the time evolution of computed $S^i$ for the stochastic parameters of the model as introduced in Table \ref{Table: stochastic_parameters}. It shows that the dominant stochastic parameter that affects the uncertainty in the kinetic energy is $\xi^3$, which represents the off-centered rotation, $\epsilon$, as we observe that $S^3 > 0.8$ at all recorded times, while the effects of the other parameters are always less than 0.2. 
In particular, by focusing on $t < 0.75$, we realize that oscillatory effect of the angular velocity model embodied in $\xi^2$, is the second dominant source of randomness propagated in the kinetic energy of the entire system, nevertheless, after $t=0.75$ as the dynamics of instability evolves with time, the effect of oscillations in the angular velocity decreases. In fact, when $0.75 < t$ the eccentric rotation is the only effective mechanism appearing in the uncertainty of kinetic energy.

On the other hand, by following the summation of the first-order sensitivity indices depicted in Figure \ref{fig: S_indc}, we observe that $\sum_i S^i > 0.95$, which reveals that the joint interactions of the stochastic parameters on the total variance of kinetic energy are negligible. However, presence of these joint interactions is slightly realized close to the onset of the instability when $t < 0.75$.

\subsection{Statistical Analysis of Fluctuating Flow Fields}\label{subsec: Fluctuations}

\begin{figure}[t!]
    \begin{minipage}[b]{.49\linewidth}
        \centering
        \includegraphics[width=1\textwidth]{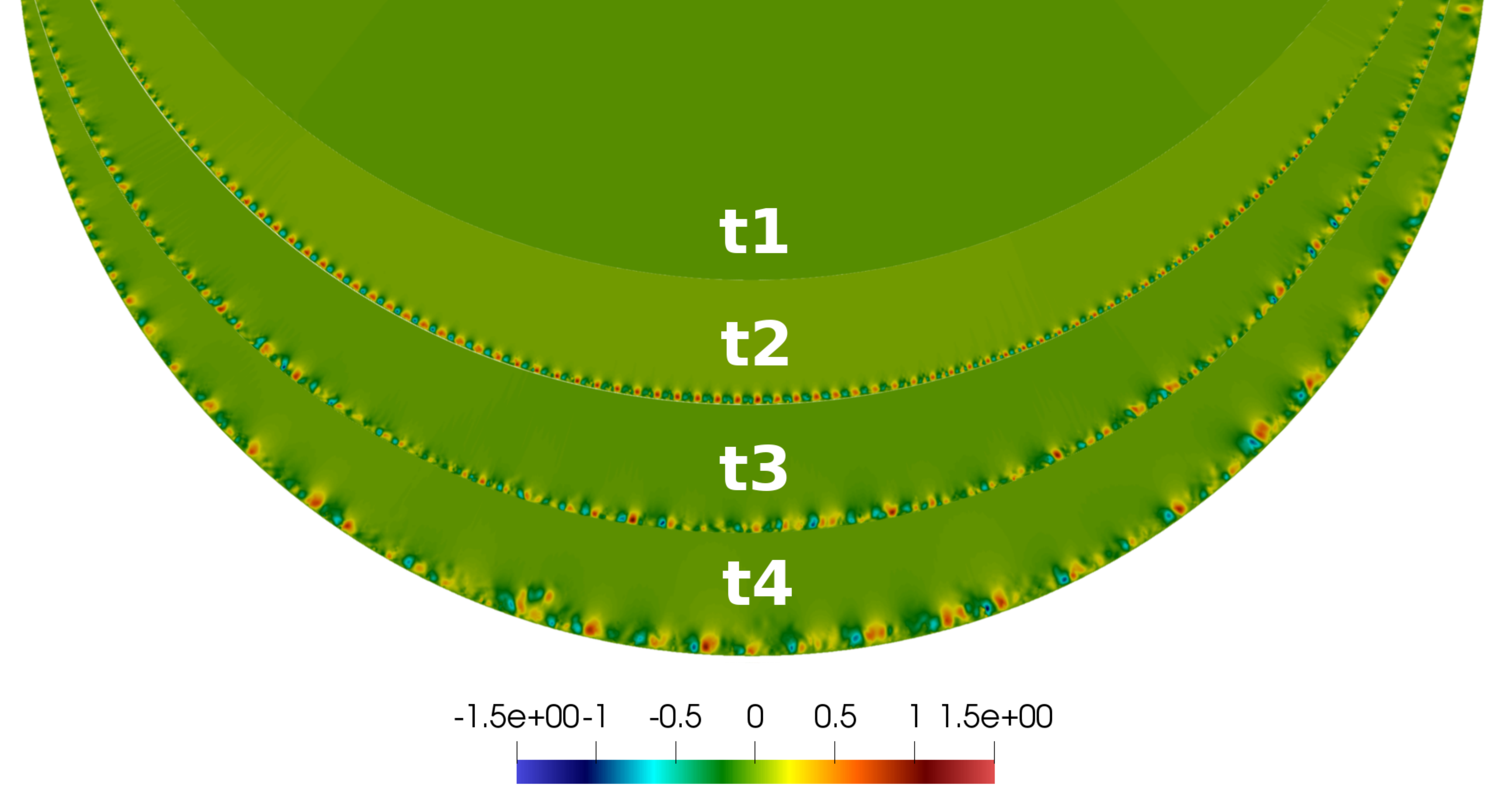}
        \subcaption{Radial velocity fluctuations, $u^\prime_r(\mathbf{x},t)$.}\label{fig: Fluctuation_vel_a}
    \end{minipage}
    \begin{minipage}[b]{.49\linewidth}
        \centering
        \includegraphics[width=1\textwidth]{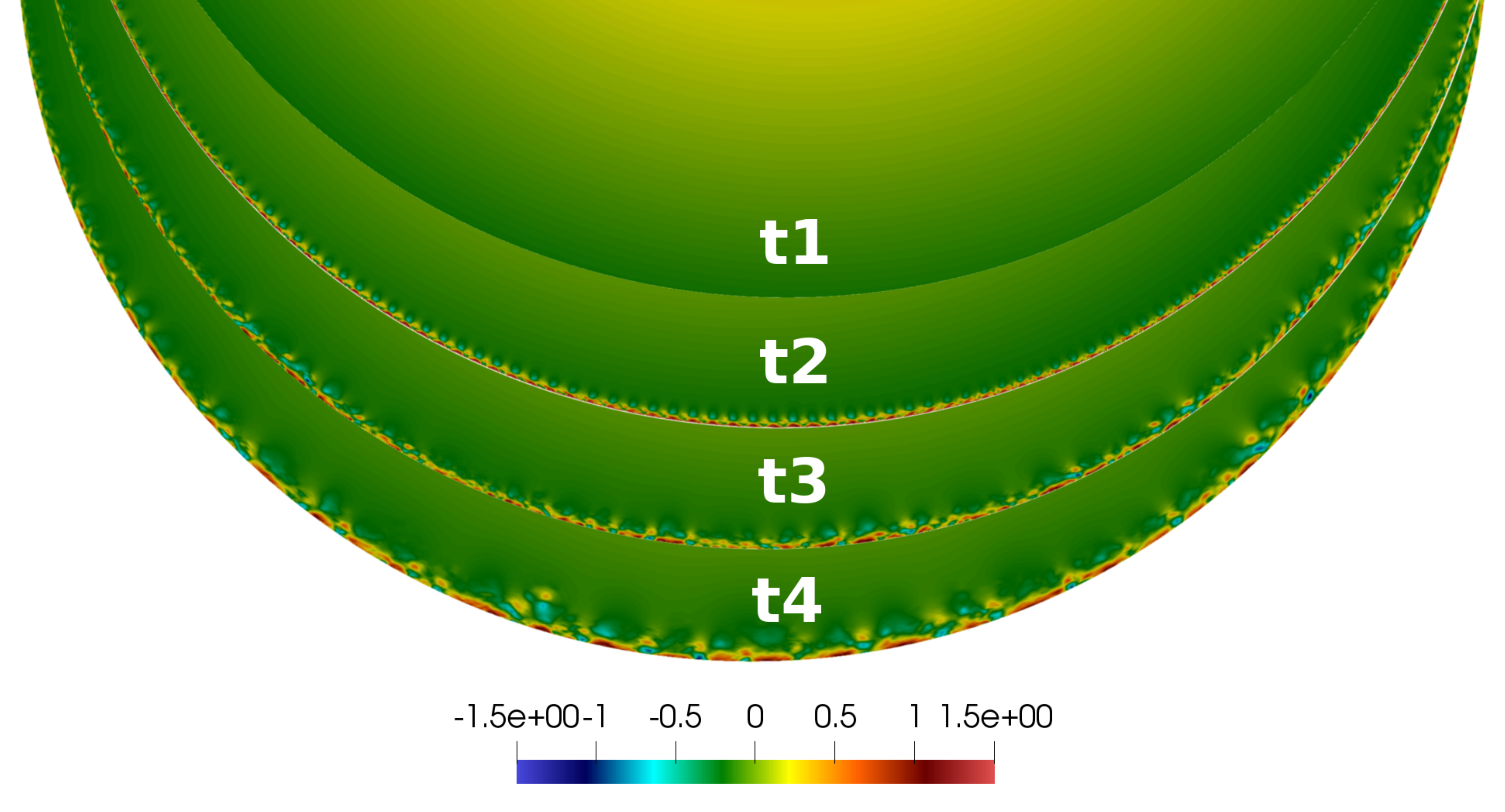}
        \subcaption{Azimuth velocity fluctuations, $u^\prime_\theta(\mathbf{x},t)$.}\label{fig: Fluctuation_vel_b}
    \end{minipage}
    \caption{Snapshots of velocity fluctuations at $t_i=0.025, \ 0.2, \ 0.375, \ 0.75$ for $i=1,\dots,4$.}\label{fig: Fluctuation_vel}
\end{figure}

Emergence of fluctuating flow velocity field plays a key role in the dynamics of flow instabilities. For instance, Ostilla \textit{et al.} \cite{Ostilla2014_velocity_PoF} studied the behavior time-averaged root-mean-square (r.m.s.) of the velocity fluctuations to study the dynamics of boundary layer in different regimes of Taylor-Couette flow. In another study, Grossmann \textit{et al.} \cite{Grossmann2014_velocity_PoF} examined the behavior of velocity fluctuations profile in a strong turbulent regime of Taylor-Couette problem. In this regard, here we seek to shed light on the mechanism of initiating the flow instability from a statistical perspective through studying the behavior of the fluctuations. In principle, any instantaneous field variable such as velocity, $\pmb{V}$, which contains a fluctuating part could be decomposed into
\begin{align}\label{eqn: vel_fluctuation}
    \pmb{V} = \big \langle \pmb{V} \big \rangle + \pmb{V}^\prime,
\end{align}
where $\pmb{V}^\prime$ represents the fluctuations of $\pmb{V}$ and $\big \langle \pmb{V} \big \rangle$ denotes its ensemble average. Unlike the applied approach in [\onlinecite{Ostilla2014_velocity_PoF, Grossmann2014_velocity_PoF}] that approximates the ensemble average by time-averaging over a time period on developed flow, here we are not allowed to exploit time-averaging close to the onset of the instability, which essentially takes place in a short period of time. However, our stochastic modeling and CFD platform enables us to properly approximate the ensemble-averaged velocity field with reasonable computational cost. Hence, having the knowledge of ensemble mean velocity field gives us the fluctuating response of the flow field variables. The fluctuations are appeared in the flow at the existence of stochasticity and disturbance in the system. In fact, the ensemble mean is nothing but finding the mathematical expectation of the field variable over the entire sample space that contains large enough number of realizations. Thus, what we obtain as the result of equation \eqref{eqn: EX} is the representation of ensemble mean in a PCM setting \cite{xiu2006numerical}. The stochastic convergence analysis we performed in section \ref{subsec: Stoch_conv_UQ} ensures that the expectation we compute from PCM is independent of the stochastic discretization, therefore, we are allowed to claim that the expected velocity field on the sufficiently converged PCM is a robust approximation of its ensemble average with large enough number of independent samples. As a result, we can write
\begin{align}\label{eqn: vel_ensembleAve}
   \big \langle \pmb{V} \big \rangle = \mathbb{E} \left[\pmb{V}(\mathbf{x},t;\pmb{\xi})\right].
\end{align}
According to the sensitivity analysis we performed in section \ref{subsec: SA_KE}, we are allowed to obtain the ensemble-averaged field by performing a uni-variate PCM on the most sensitive stochastic parameter, $\xi^3=\epsilon$, while we fix the other two random parameters of the wall velocity model to their mean values as reported in the Table \ref{Table: stochastic_parameters}. Since the uni-variate PCM requires much less realizations evaluated at collocation points, it is computationally feasible to discretize the dominant random direction even beyond the stochastic convergence resolution. Here we proceed with taking 30 collocation/integration points providing a high-resolution expected solution in the stochastic space essentially returning a seamless evaluation of $\big \langle \pmb{V} \big \rangle$. According to Table \ref{Table: stochastic_parameters} and as a physically reasonable assumption, the rotational eccentricity is initially taken to be varying up to 5\% of the cylinder radius as $\epsilon \sim \mathcal{U}\left( 0.0,0.05\right)$. For a randomly drawn realization of the sample space that fixes eccentricity value at $\epsilon=0.0263$, we evaluate the fluctuating velocity field according to equation \eqref{eqn: vel_fluctuation}. The procedure of computing the fluctuations from SEM-based realizations is briefly explained in \ref{sec: Appendix2}. Figure \ref{fig: Fluctuation_vel} shows the resulting velocity fluctuations in polar coordinate system at four snapshots of time illustrating the onset of flow instability. 

\subsubsection{Emergence of Non-Gaussian Statistics in Velocity Fluctuations}\label{subsec: vel_Fluctuations}

\begin{figure}[t!]
    \begin{minipage}[b]{.49\linewidth}
        \centering
        \includegraphics[width=1\textwidth]{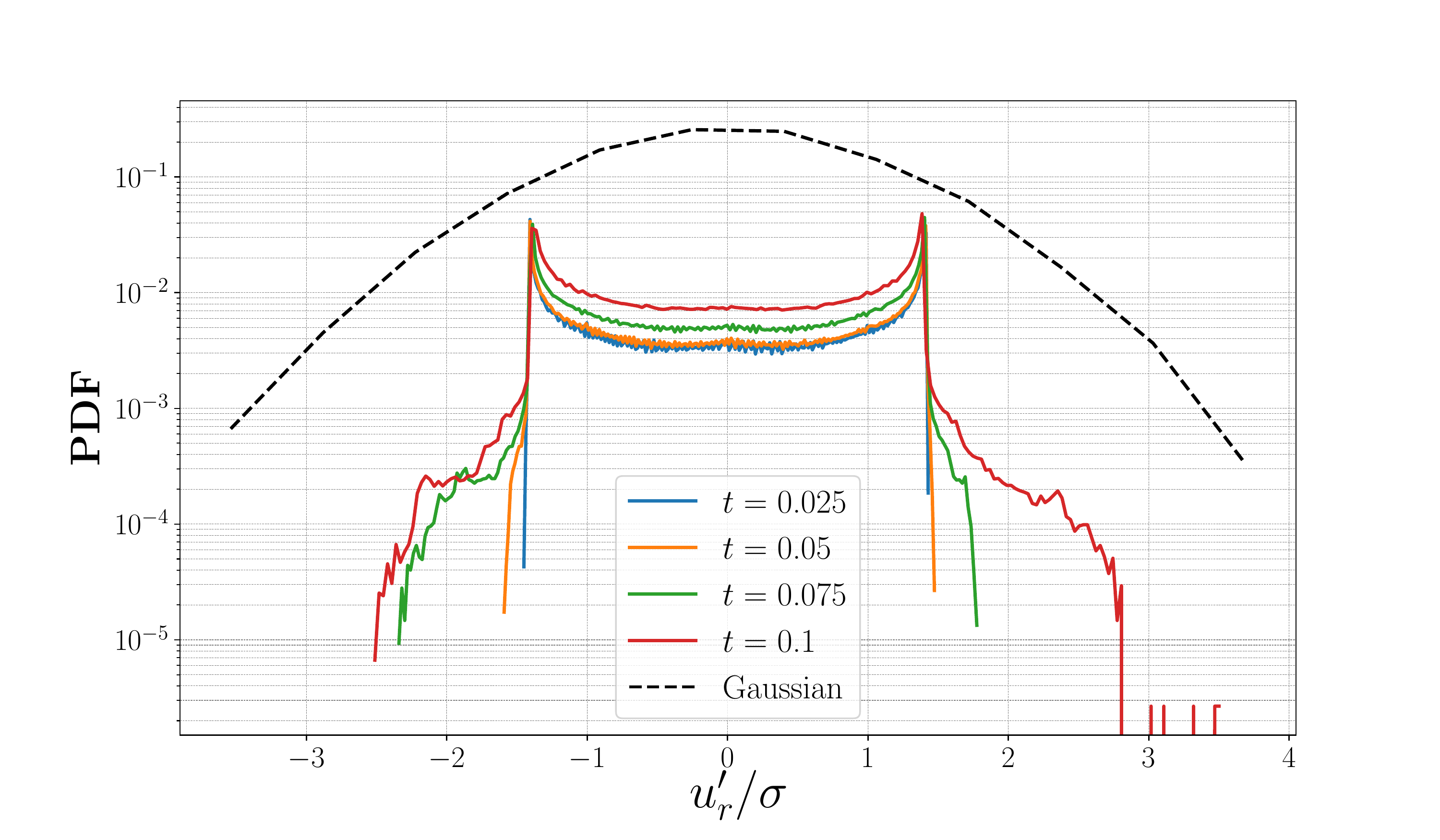}
        \subcaption{Radial velocity fluctuations, $0 < t \leq 0.1$}\label{fig: PDF_ur_a}
    \end{minipage}
    \begin{minipage}[b]{.49\linewidth}
        \centering
        \includegraphics[width=1\textwidth]{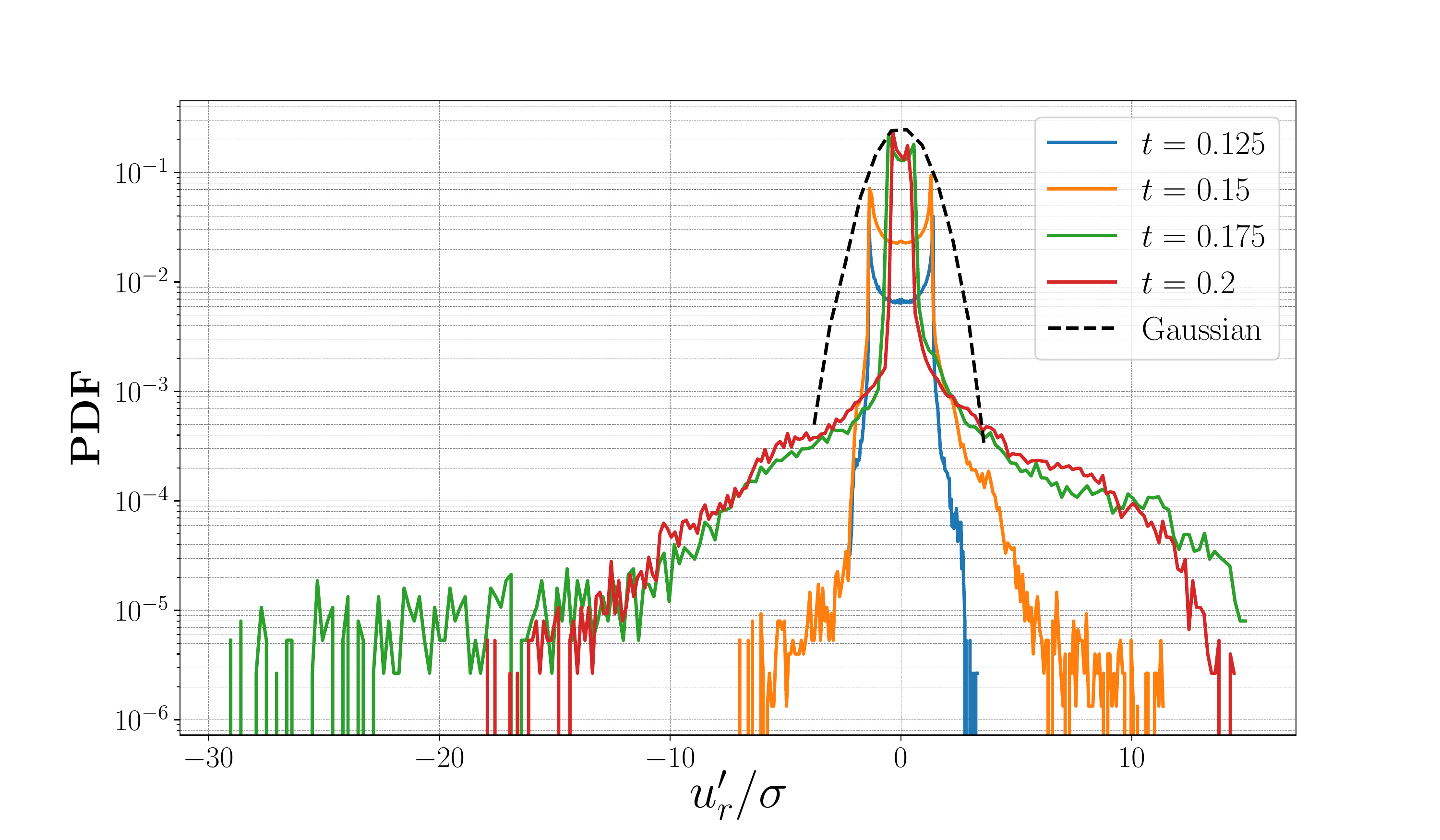}
        \subcaption{Radial velocity fluctuations, $0.1 < t \leq 0.2$}\label{fig: PDF_ur_b}
    \end{minipage}
    \begin{minipage}[b]{.49\linewidth}
        \centering
        \includegraphics[width=1\textwidth]{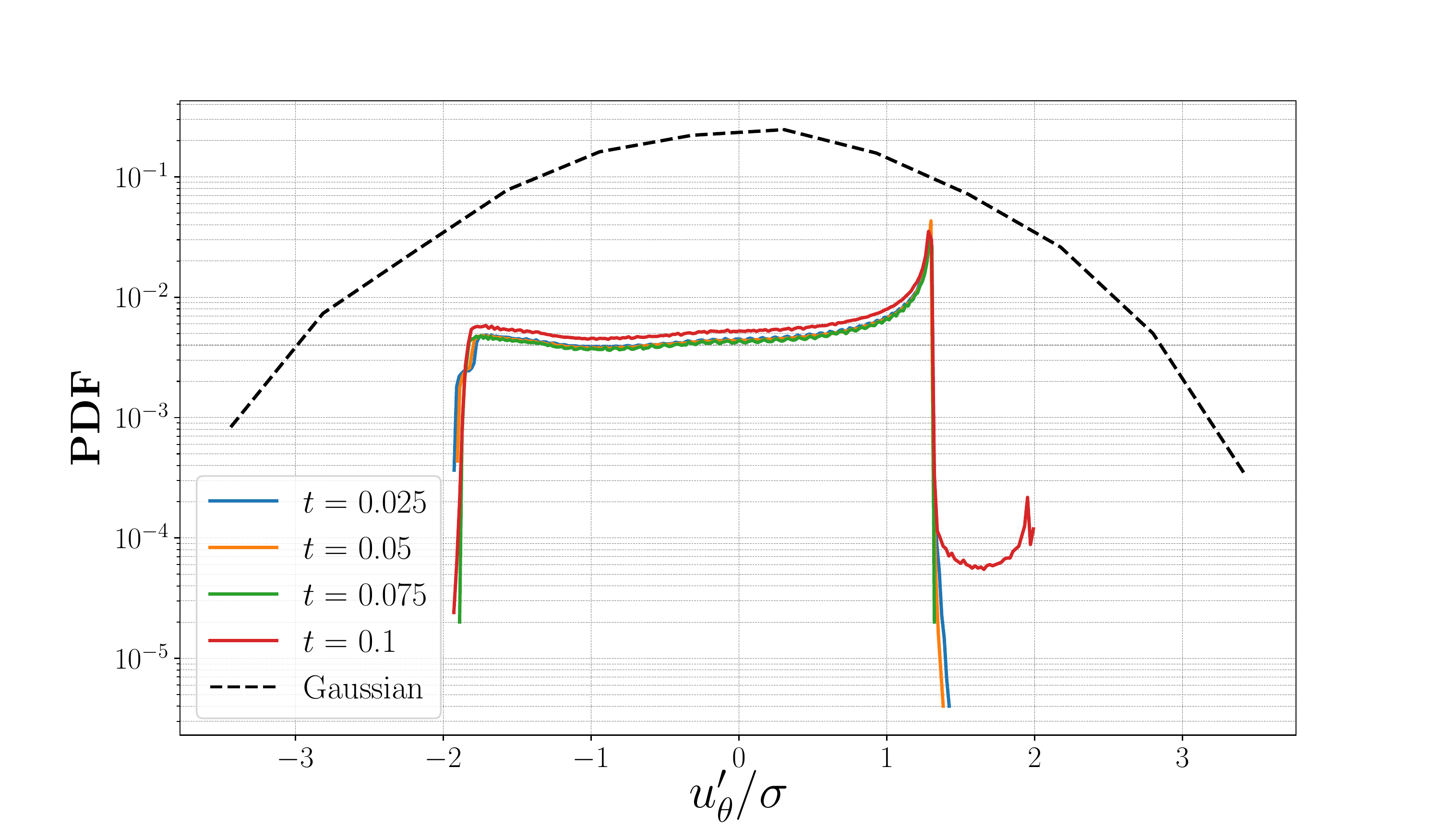}
        \subcaption{Azimuth velocity fluctuations, $0 < t \leq 0.1$}\label{fig: PDF_ut_a}
    \end{minipage}
    \begin{minipage}[b]{.49\linewidth}
        \centering
        \includegraphics[width=1\textwidth]{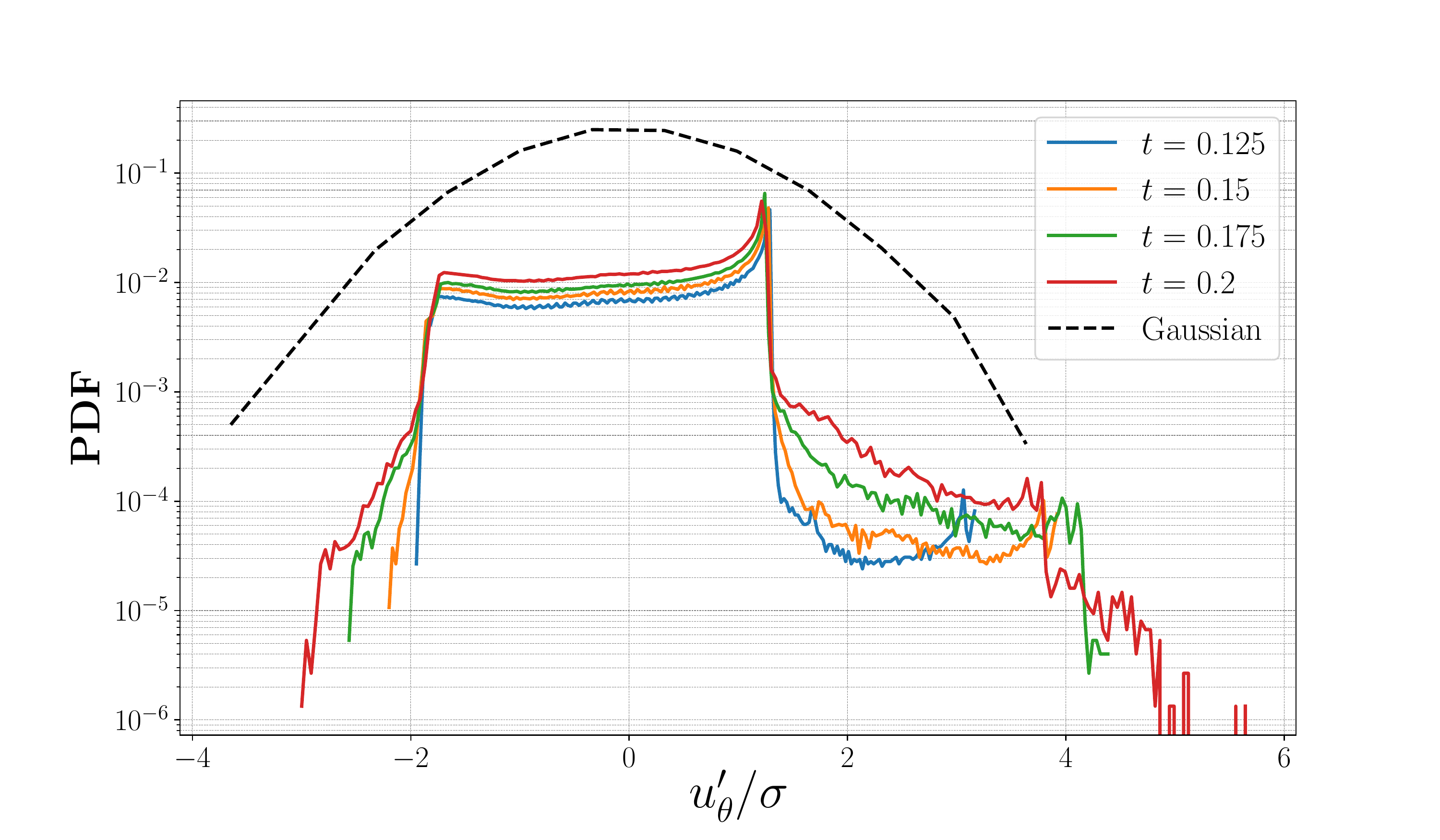}
        \subcaption{Azimuth velocity fluctuations, $0.1 < t \leq 0.2$}\label{fig: PDF_ut_b}
    \end{minipage}
    \caption{Time evolution of PDFs of components of the velocity fluctuations for eight instances of time close to the flow instability onset. Here all the PDFs are obtained for the fluctuations normalized by their own standard deviations, $\sigma$, and they are all compared with the standard Gaussian PDF, $\mathcal{N}(0,1)$.}\label{fig: PDFs_u}
\end{figure}

\begin{figure}[t!]
    \begin{minipage}[b]{.49\linewidth}
        \centering
        \includegraphics[width=1\textwidth]{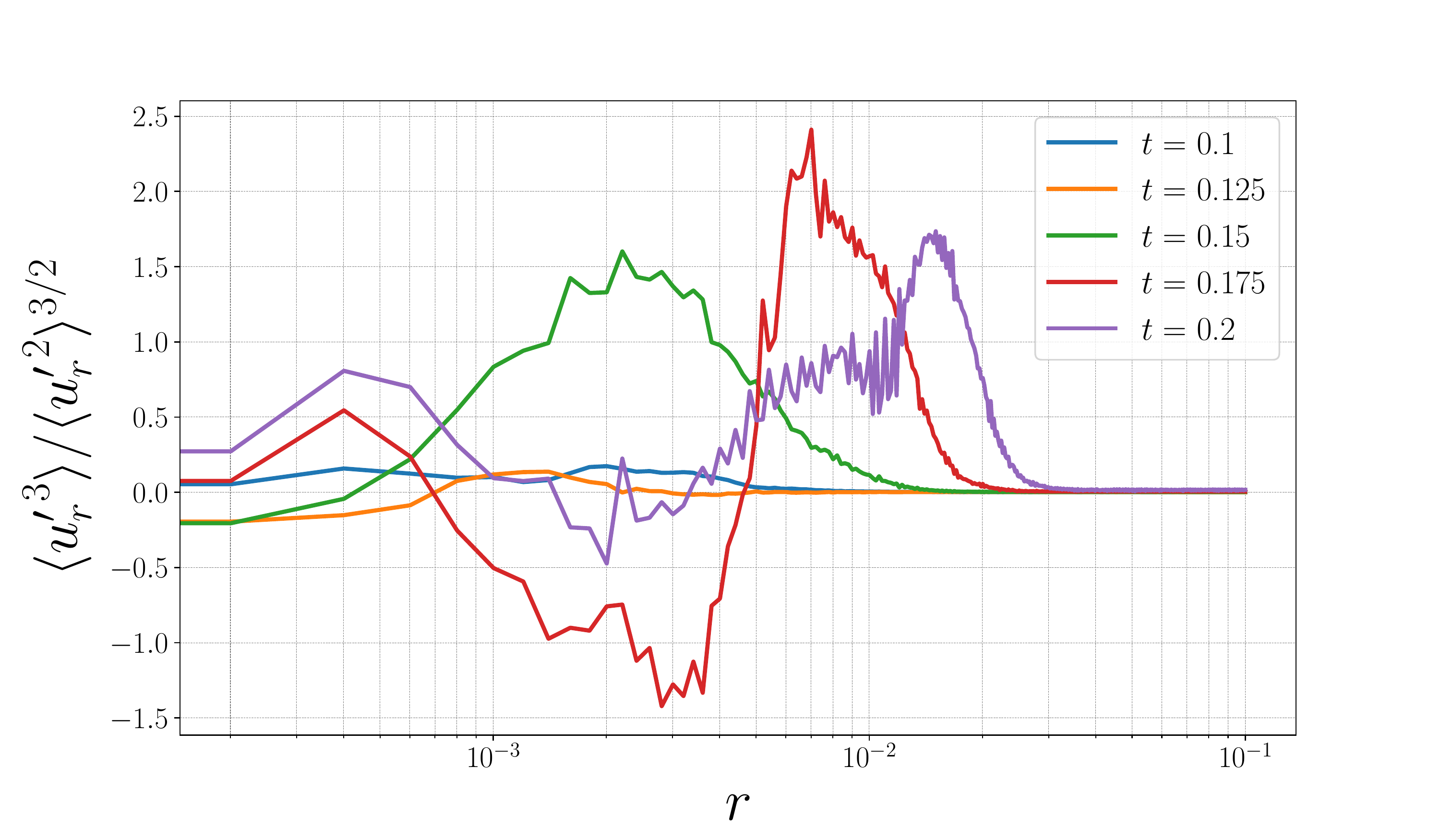}
        \subcaption{Skewness factor for $u^\prime_r$}\label{fig: ur_Skw}
    \end{minipage}
    \begin{minipage}[b]{.49\linewidth}
        \centering
        \includegraphics[width=1\textwidth]{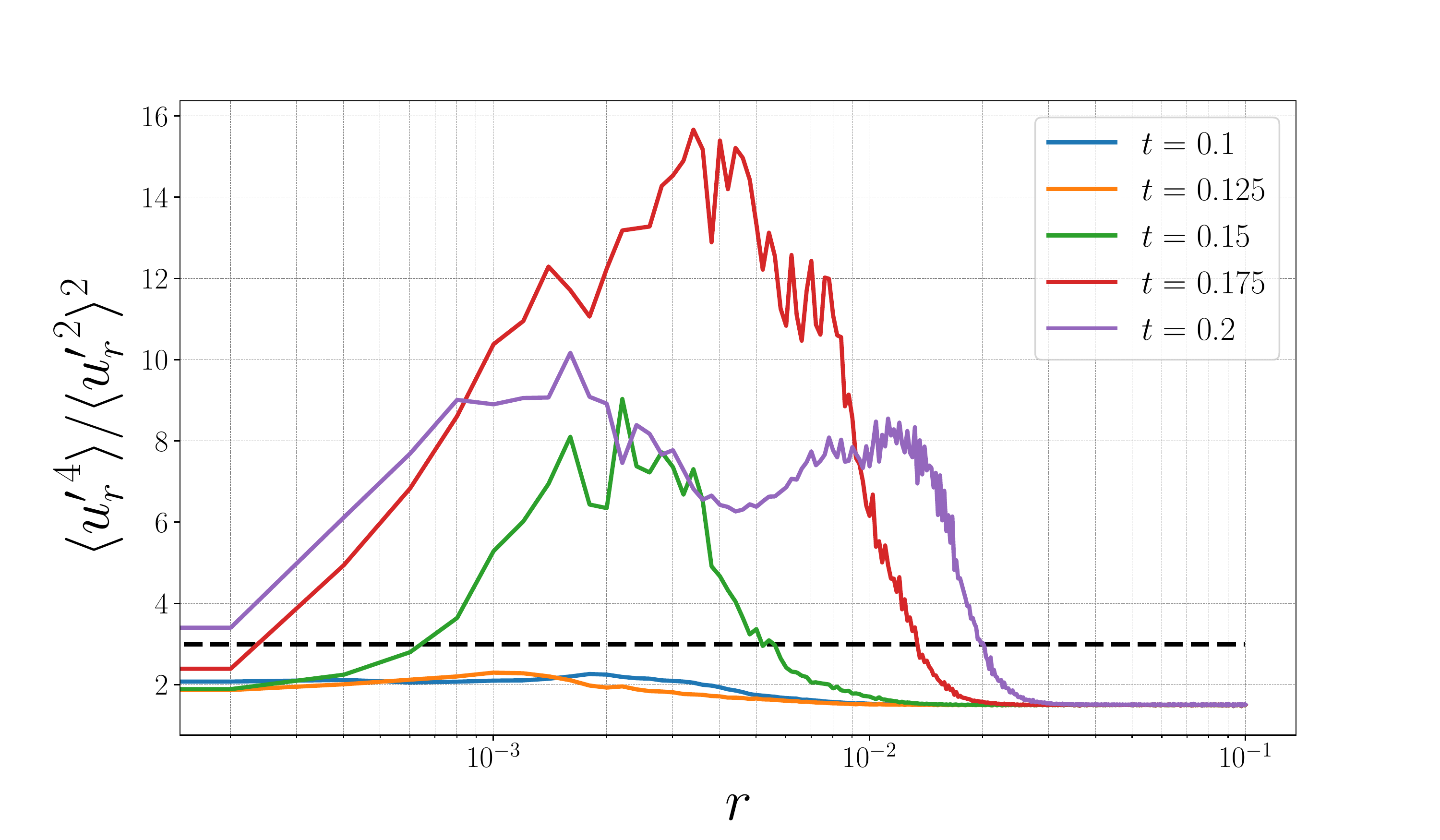}
        \subcaption{Flatness factor for $u^\prime_r$}\label{fig: ur_Flt}
    \end{minipage}
    
    \begin{minipage}[b]{.49\linewidth}
        \centering
        \includegraphics[width=1\textwidth]{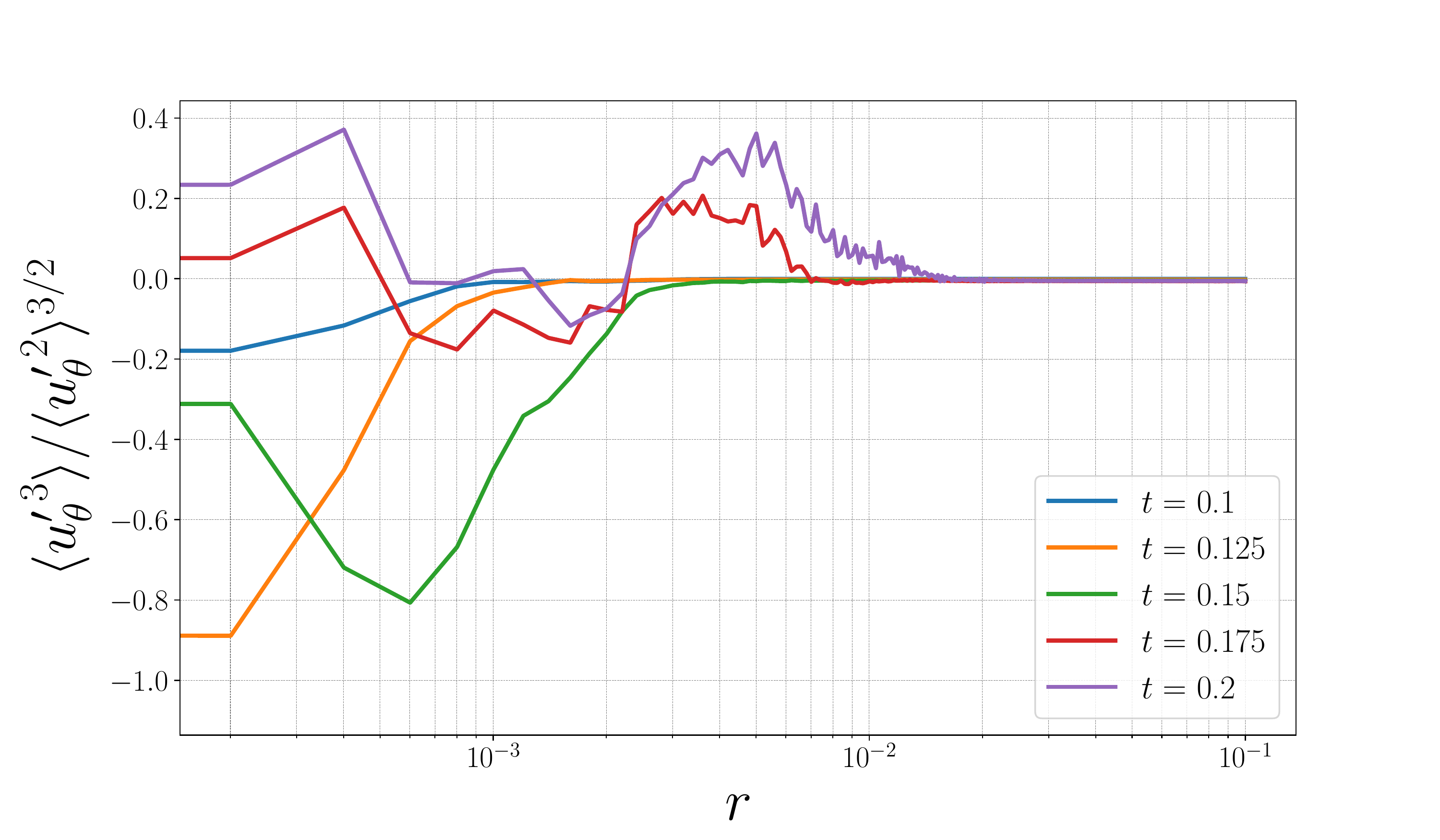}
        \subcaption{Skewness factor for $u^\prime_\theta$}\label{fig: ut_Skw}
    \end{minipage}
    \begin{minipage}[b]{.49\linewidth}
        \centering
        \includegraphics[width=1\textwidth]{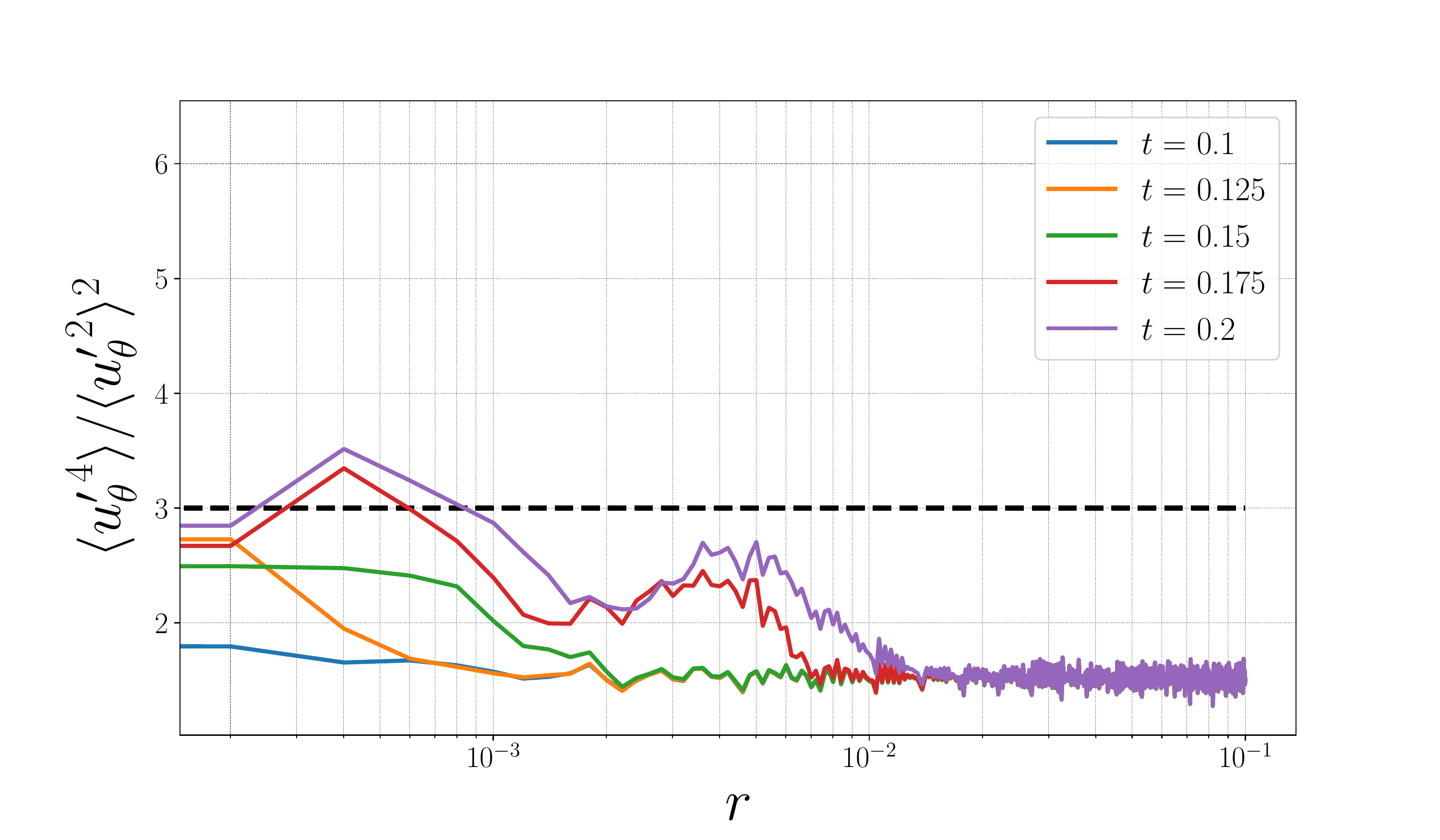}
        \subcaption{Flatness factor for $u^\prime_\theta$}\label{fig: ut_Flt}
    \end{minipage}
    \caption{High-order moments of velocity fluctuations, $\pmb{V}^\prime = (u^\prime_r,u^\prime_\theta)$, as a function of radial distance from the wall, $r$, where $r=0$ indicates the wall. In Figures \ref{fig: ur_Flt} and \ref{fig: ut_Flt}, the black-colored dashed lines indicate the flatness factor associated with the standard Gaussian distribution.}\label{fig: Fluctuation_High_Moments}
\end{figure}

\begin{figure}[t!]
    \begin{minipage}[b]{.49\linewidth}
        \centering
        \includegraphics[width=1\textwidth]{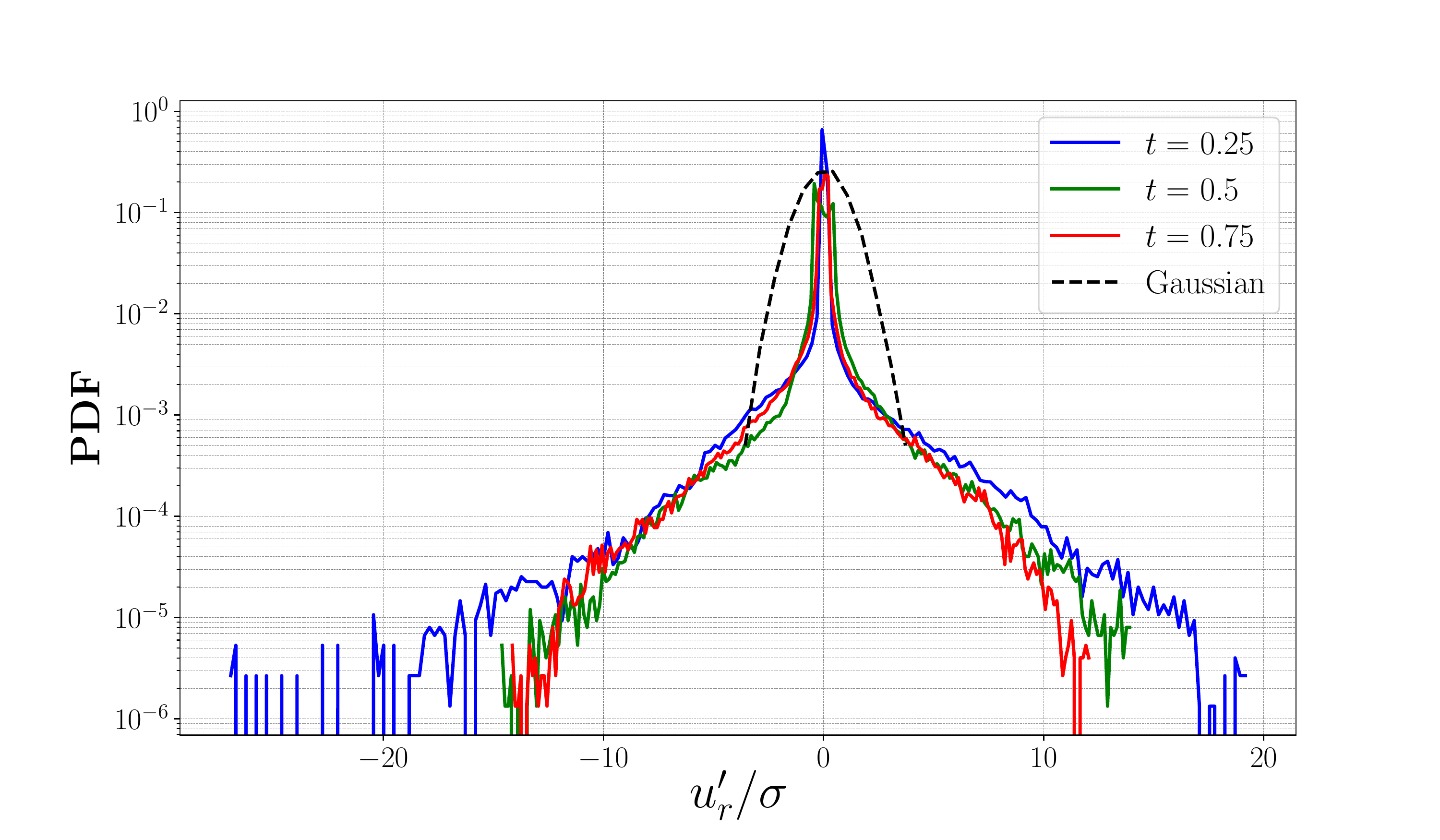}
        \subcaption{Radial velocity fluctuations.}\label{fig: PDFs_longer_time-ur}
    \end{minipage}
    \begin{minipage}[b]{.49\linewidth}
        \centering
        \includegraphics[width=1\textwidth]{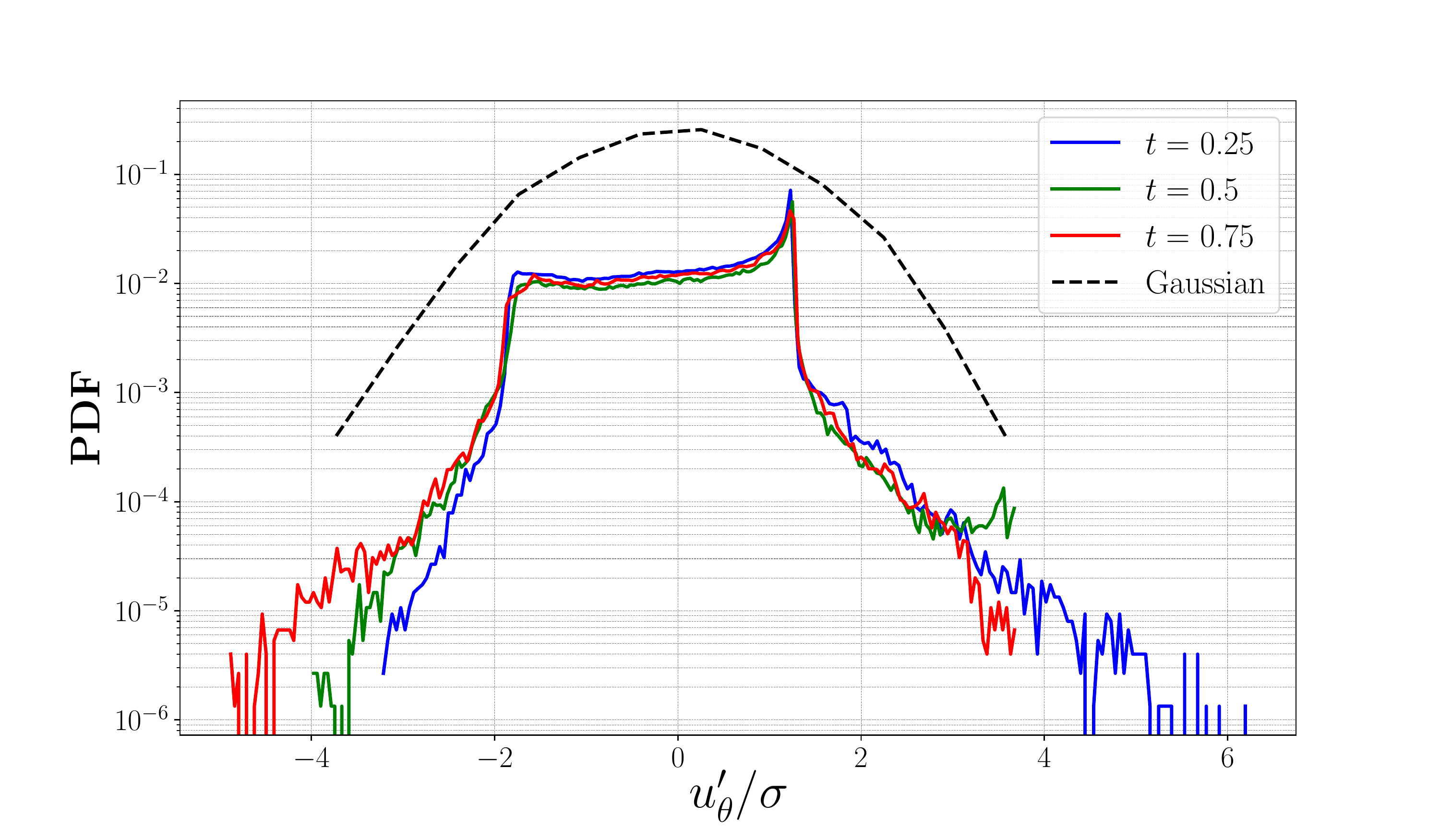}
        \subcaption{Azimuth velocity fluctuations.}\label{fig: PDFs_longer_time-ut}
    \end{minipage}
    \caption{Comparison between the standard Gaussian PDF and PDFs of the velocity fluctuations at $t=0.25, \ 0.5, \ 0.75$.}\label{fig: PDFs_longer_time}
\end{figure}

Tracking the probability density function (PDF) of velocity fluctuations with time renders qualitative statistical information, which characterizes the impacts of the evolution of fluctuations on the dynamics. PDF of fluctuating fields can simply show us the departure from Gaussian statistical behavior that essentially plays an important role in leading to a chaotic flow dynamic state. Here, we compute the velocity fluctuations' PDFs over the computational domain for the radial and azimuth components, and plot them at eight different time states close to the initiation of the flow instability (see Figure \ref{fig: PDFs_u}). All of these PDFs are computed for the velocity fluctuations that are normalized by their standard deviation so that the comparison with the standard Gaussian PDF, drawn from $\mathcal{N}(0,1)$, is readily possible through eyeball measure. Here, Figures \ref{fig: PDF_ur_a} and \ref{fig: PDF_ut_a} are depicting the PDFs of normalized radial and azimuth components of velocity fluctuations for $0 < t \leq 0.1$, respectively. For both of the radial and azimuth velocity components the PDFs are showing sub-Gaussian behavior that is commonly expected given the laminar initial state of the flow, however, the former rapidly tends to show broader tails compared to the latter with time. Moreover, we can observe that the onset of the flow instability causes noticeable deviations from symmetry in the PDF of radial velocity fluctuations. By tracking the PDFs of velocity fluctuations at further times, \textit{i.e.} $0.1 < t \leq 0.2$, one can clearly observe that emergence of broad PDF tails and asymmetries quickly lead to a highly non-Gaussian statistical behavior (see Figures \ref{fig: PDF_ur_b} and \ref{fig: PDF_ut_b} and compare with the standard Gaussian PDF). More specifically, Figure \ref{fig: PDF_ur_b} shows that the velocity fluctuations in the radial direction are essentially the main source of this non-Gaussianity as the heavy-tailed PDF accompanied with intermittent events distributed at the PDF tails are arising (see $0.15 \leq t \leq 0.2$). On the other hand, a noticeable skewness towards the negative-valued fluctuations of the radial velocity component tends to grow with time as shown in Figure \ref{fig: PDF_ur_b}. Comparing the radial and azimuth components of velocity fluctuations qualitatively show that emerging the aforementioned features that are essentially the fingerprints of non-Gaussian statistics is much milder and at slower rates for the azimuth component, $u^\prime_\theta$.

In order to obtain a quantitative measure on the non-Gaussian statistics of the velocity fluctuations, we manage to compute their skewness and flatness factors as a function of radial distance from the wall, $r$. This effectively helps to understand how the non-Gaussian behavior evolves through time as we move away from the wall towards the center. Our approach involves uniformly sampling the velocity values on the circular stripes with a thickness of $\delta r$ where their radial distance from the wall is $r$. Once we performed such sampling, we can simply attain the skewness and flatness factors as $\langle {\pmb{V}^\prime}^3 \rangle/\langle {\pmb{V}^\prime}^2 \rangle^{3/2}$ and $\langle {\pmb{V}^\prime}^4 \rangle/\langle {\pmb{V}^\prime}^2 \rangle^2$, respectively. In our measurements, we took $\delta r = 2\times 10^{-4}$ and $\langle \cdot \rangle$ denotes spatial averaging over the uniformly sampled velocity space on each circular stripe with radial distance $r$ from wall. As a result, Figure \ref{fig: Fluctuation_High_Moments} illustrates such radial skewness and flatness factors for both components of velocity fluctuations at five instances of time for $0.1 \leq t \leq 0.2$. The resulting measures for $u^\prime_r$ depicted in Figures \ref{fig: ur_Skw} and \ref{fig: ur_Flt} show that the non-zero skewness factor and flatness factor greater than 3 (measures associated with standard Gaussian) are appearing for $0.15 \leq t$. This record is in total agreement with what we observe in their non-Gaussian PDFs in Figure \ref{fig: PDF_ur_b}. For $u^\prime_\theta$, Figure \ref{fig: ur_Skw} illustrates non-zero skewness factor values close to the wall at all the recorded times and Figure \ref{fig: ur_Flt} shows that for a narrow region close to the wall the flatness factor exceeds 3 for $0.15 < t$. Again, these observations are in complete agreement with the behavior we observe in PDFs of $u^\prime_\theta$ shown in Figure \ref{fig: PDF_ut_b}. More specifically on the heavy-tailed velocity fluctuations PDFs, one can link the radial records of flatness factor in both components $u^\prime_r$ and $u^\prime_\theta$ as shown in Figures \ref{fig: ur_Flt} and \ref{fig: ut_Flt}, respectively. In radial velocity fluctuations, it is clearly seen that as time passes the flatness factor increases for the closest radial distances to wall, \textit{i.e.} $r < 10^{-3}$, and in farther distances from the wall, a span of radial region of high flatness factor that essentially contributes to the rare events occurring at the PDF tails (for $0.15 \leq t$) is observed. As we pointed out, this high flatness factor span is expanding towards the center of the cylinder as flow instability evolves in time. Although such behavior is also seen for the azimuth component of velocity fluctuations, its intensity is much milder compared to $u^\prime_r$. In fact, our records show that for $u^\prime_\theta$ the flatness factor rarely exceeds 3 (see Figure \ref{fig: ut_Flt}).

Finally, by comparing the PDFs of velocity fluctuations for $0.2<t$ with the one associated with standard Gaussian (see Figure \ref{fig: PDFs_longer_time}), we recognize that the statistical features such as non-symmetric distributions and heavy PDF tails with high intermittency are remarkably discernible. However, as illustrated for the prior times closer to the flow instability initiation, these features seem to be manifested more prominently in the radial component of velocity fluctuations.

\subsubsection{Memory Effects in Vorticity Dynamics and Anomalous Time-Scaling of Enstrophy}\label{subsec: vor_Fluctuations}

\begin{figure}[t!]
    \begin{minipage}[b]{1\linewidth}
        \centering
        \includegraphics[width=.8\textwidth]{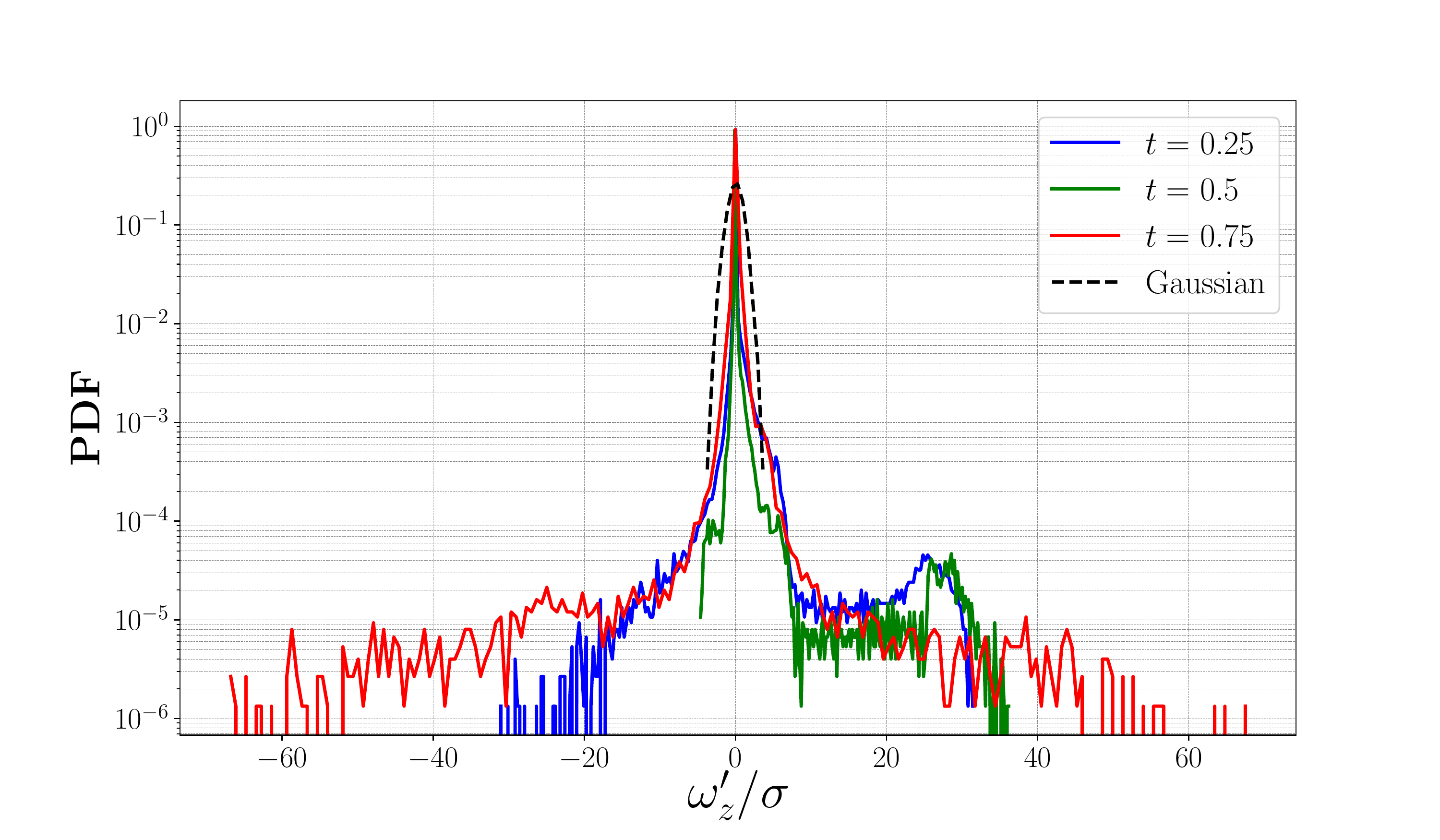}
        \subcaption{Comparison between the standard Gaussian PDF and normalized vorticity fluctuations' PDFs at $t=0.25, \ 0.5, \ 0.75$.}\label{fig: PDF_longer_time_wz}
    \end{minipage}
    
    \begin{minipage}[b]{.49\linewidth}
        \centering
        \includegraphics[width=1\textwidth]{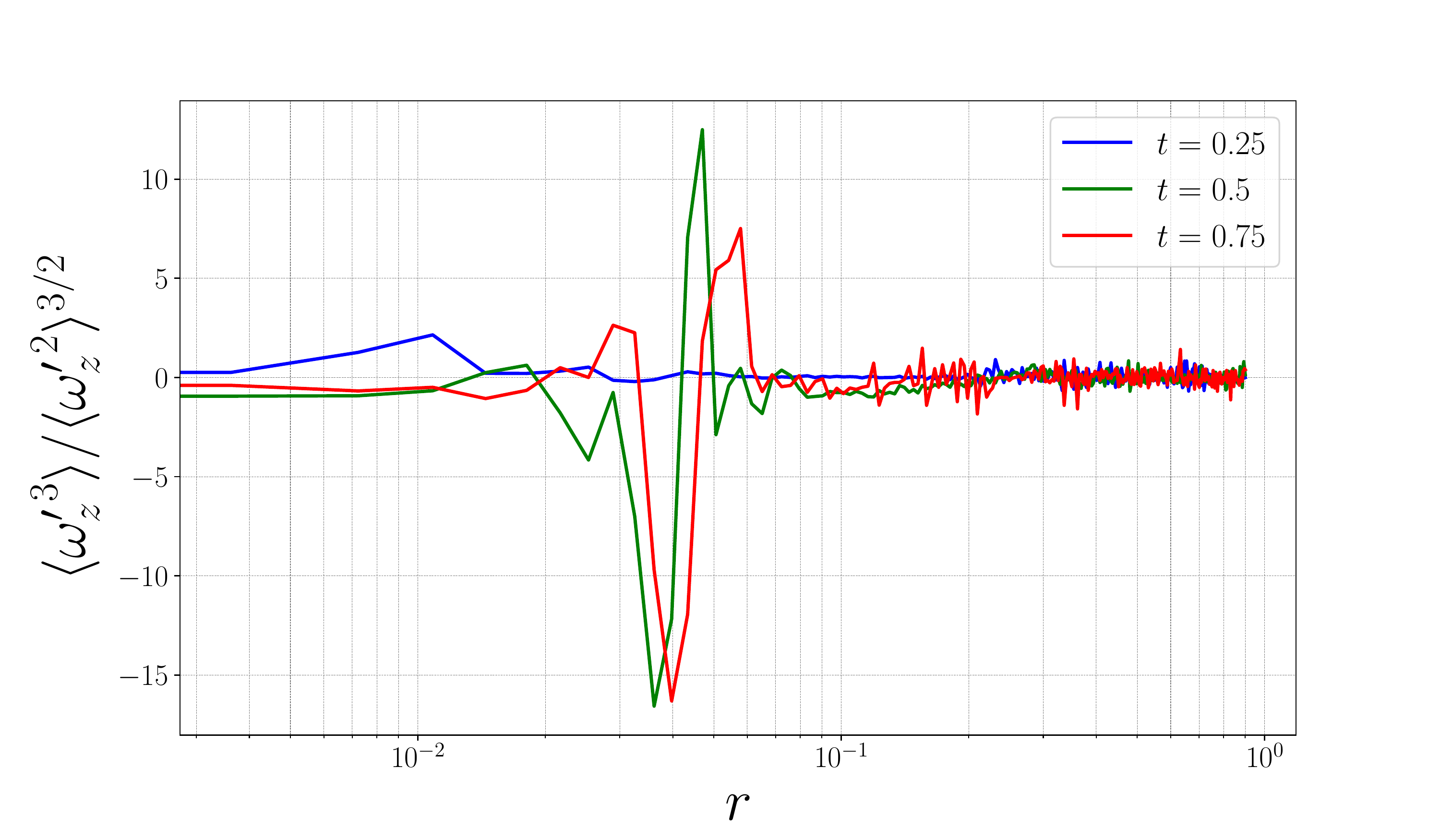}
        \subcaption{Skewness factor for $\omega^\prime_z$.}\label{fig: Skw_wz}
    \end{minipage}
    \begin{minipage}[b]{.49\linewidth}
        \centering
        \includegraphics[width=1\textwidth]{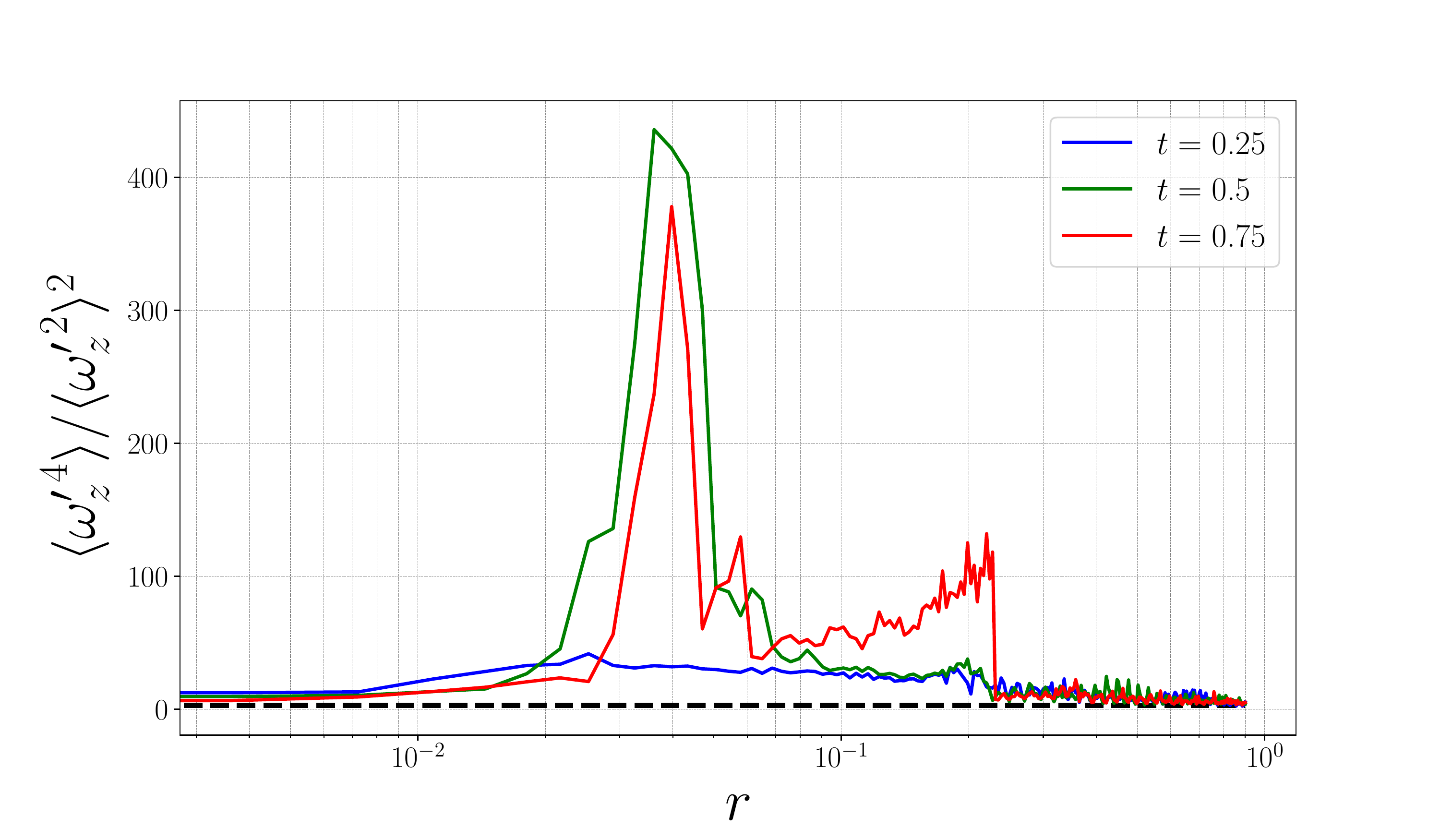}
        \subcaption{Flatness factor for $\omega^\prime_z$.}\label{fig: Flt_wz}
    \end{minipage}
    \caption{Comparison between the PDFs of vorticity fluctuations at $t=0.25, \ 0.5, \ 0.75$ and standard Gaussian PDF. Dashed lines indicate the measures associated with Gaussian behavior.}\label{fig: longer_time_wz}
\end{figure}

Although early theories of Batchelor \cite{batchelor1969computation} assumed that for decaying two-dimensional turbulence it is only kinetic energy that is mainly remembered for a long time, later it has been shown that vorticity field plays a key role in the flow dynamics, which was initially failed to be addressed by Batchelor \cite{benzi1988self}. Here, while the filamentation of the vorticity field is occurring, there exist small yet sufficiently strong patches of vorticity surviving the filamentation process and comprise coherent vortices that somehow live even longer than many large-eddy turnover times \cite{davidson2015turbulence}.  These coherent vortices are interacting with each other quite similar to a collection of point vortices. On some occasions, these coherent vortices could approach each other and merge into larger ones. Therefore, the number of coherent vortices decreases while their average size increases as flow evolves. On the other hand, given the discussion on non-Gaussian behavior velocity fluctuations, one can make a connection between the statistical behavior of the vorticity field and generation and intensity of coherent vortices resulting from the flow instability. Thus, similar to the procedure in the previous section, we compute the vorticity PDFs in addition to the radial skewness and flatness factors for the same realization of the fluctuating flow field we considered. Figure \ref{fig: longer_time_wz} provides this statistical information at $t = 0.25, 0.5,$ and 0.75. Comparing the vorticity PDFs shown in Figure \ref{fig: PDF_longer_time_wz} to the standard Gaussian PDF makes it evident that fingerprints of non-Gaussian statistics, \textit{i.e.} non-symmetric probability distributions in addition to broad and intermittent PDF tails, are immensely evolving in vorticity field. Moreover, the radial skewness and flatness factors obtained for these three time instances quantitatively demonstrate that such intense non-Gaussian statistical behavior is swiftly extending towards the center of cylinder (see the radial region of $0.01<r<0.2$ at Figures \ref{fig: Skw_wz} and \ref{fig: Flt_wz}).

\begin{figure}[t!]
    \begin{minipage}[b]{1\linewidth}
    \centering
    \includegraphics[width=.9\textwidth]{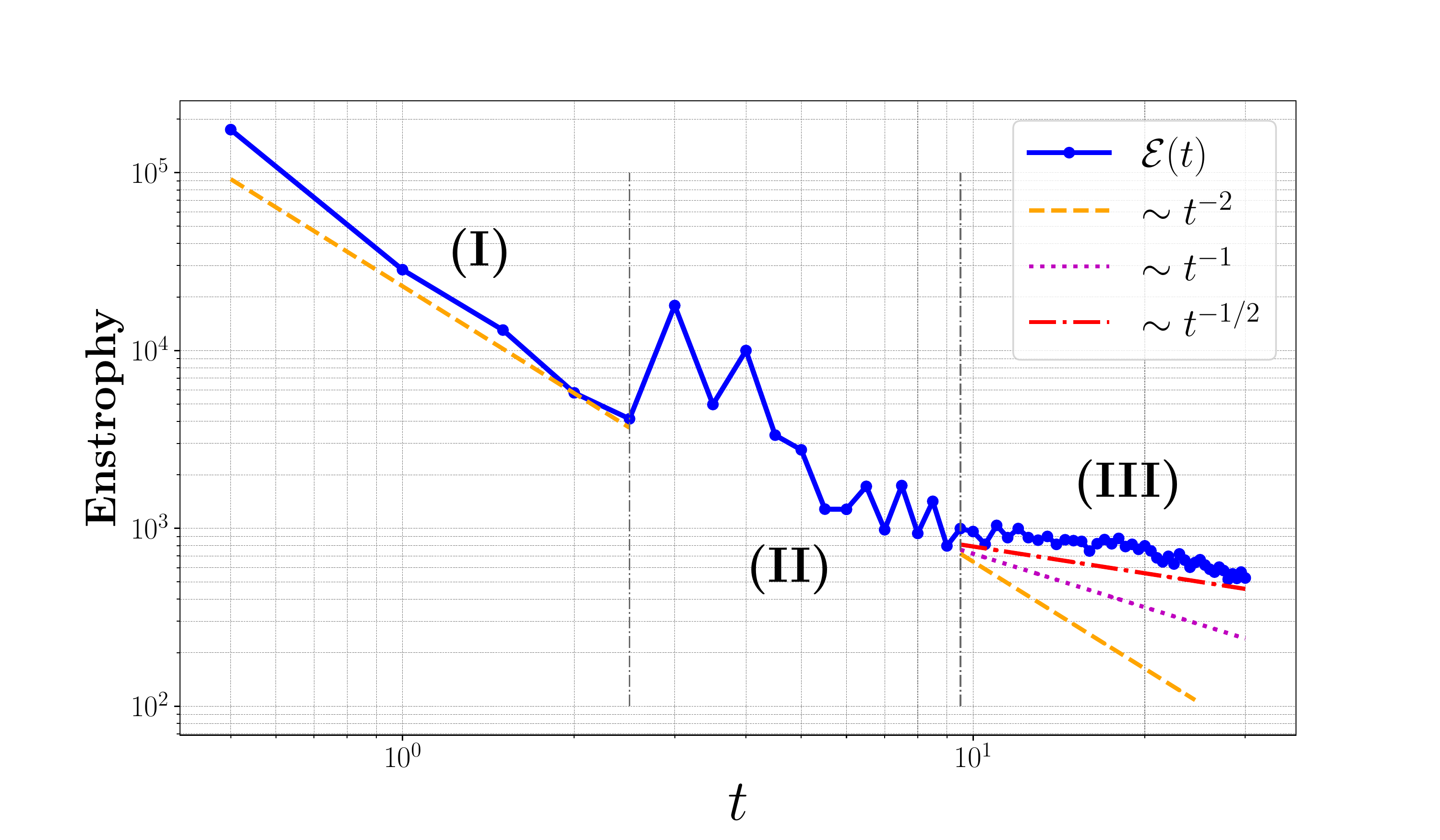}
    \subcaption{Enstrophy record, $\mathcal{E}(t)$, and its early-time \textbf{(I)}, transient-time \textbf{(II)}, and long-time \textbf{(III)} scaling affected by symmetry-breaking disturbances imposed on the rotational motion of cylinder.}\label{fig: Scaling_enstrophy}
    \end{minipage}
    
    \vspace{0.2in}
    \begin{minipage}[b]{1\linewidth}
        \begin{minipage}[b]{.49\linewidth}
            \centering
            \includegraphics[width=1\textwidth]{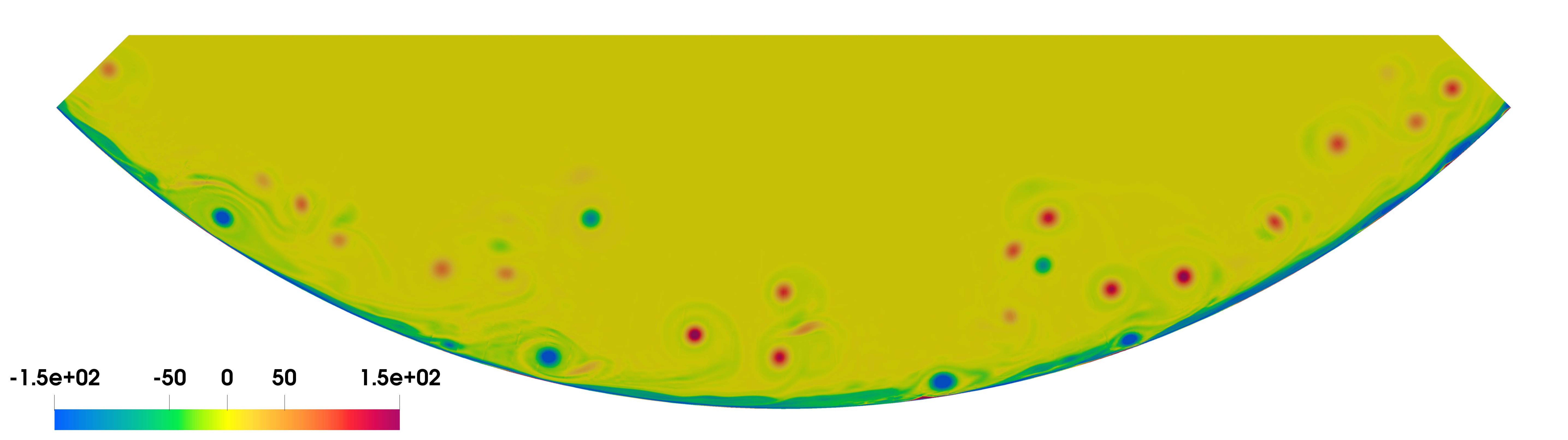}
            \caption*{$\omega_z(\mathbf{x},t=7)$}
        \end{minipage}
        \begin{minipage}[b]{.49\linewidth}
            \centering
            \includegraphics[width=1\textwidth]{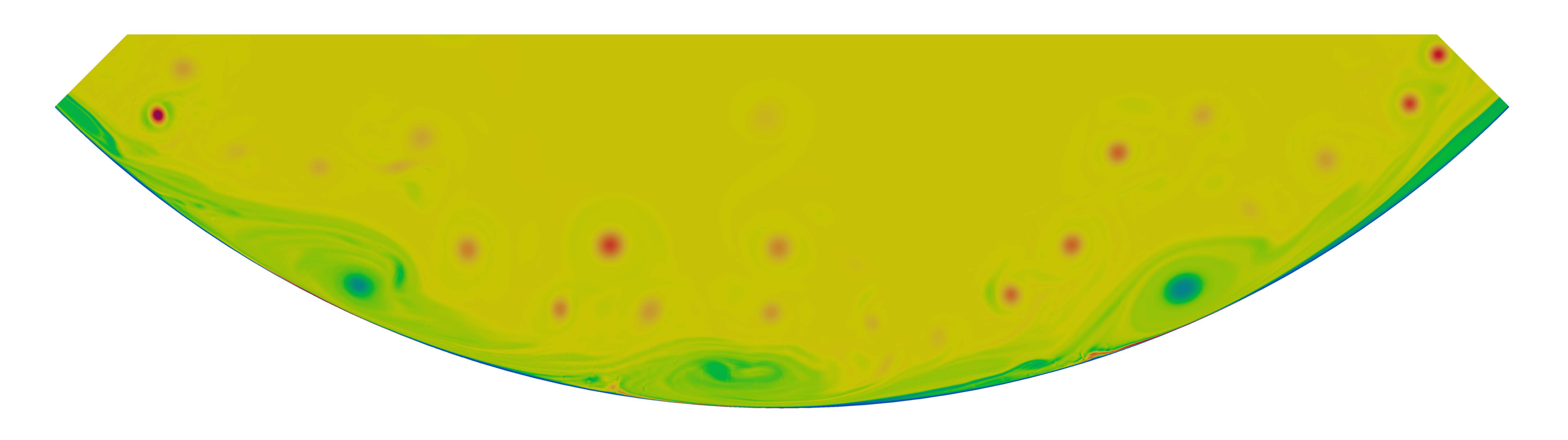}
            \caption*{$\omega_z(\mathbf{x},t=24)$}
        \end{minipage}
        \subcaption{Snapshots of instantaneous vorticity field, $\omega_z(\mathbf{x},t)$, showing the structure and growth of coherent vortical regions attached to the cylinder's wall.}\label{fig: wz_coherent}
    \end{minipage}
    \caption{Time-scaling of enstrophy record and its link to evolution of coherent vortical structures.}\label{fig: coherent_str}
\end{figure}

Given the discussion on the generation and evolution of the coherent vortices, and our quantitative/qualitative study on the emergence of strong non-Gaussian statistical behavior for velocity and vorticity fluctuations, one can argue that such statistics are closely tied to and in other words, the direct result of generation and growth of coherent vortical structures due to the effect of the rotational symmetry-breaking factors. In prior studies, such connection was investigated and partially addressed in the contexts of planar mixing and free shear layers \cite{Zayernouri2011, fathali2016sensitivity, dong2020coherent}, subgrid-scale (SGS) motions and their nonlocal modeling for homogeneous and wall-bounded turbulent flows \cite{samiee2020_FSGS, diLeoni2020_two-point}, boundary layer flows \cite{yang2020scaling, atzori2020coherent}, and turbulent flows interacting with wavy-like moving/actuated surfaces (with application to reduction and control of flow separation) \cite{akbarzadeh2019_PoF, akbarzadeh2020controlling}.

Here, an interesting yet, practical question that could be raised is that if such ``intensified'' coherent vortical structures induced by the symmetry-breaking parameters in the rotational motion are capable of incorporating more memory effects into the dynamics of vorticity field. This potentially could lead to the engineering means to increase effective chaotic mixing in rotating systems by introducing factors that initiate deviation from symmetry in rotation. In a two-dimensional turbulent/chaotic flow, the very presence of ``long-lived'' coherent vortices normally cause the time-scaling of enstrophy record at long-time to be close to $t^{-1}$, however, it initially is scaled with $t^{-2}$ at the early stages of flow which is also what Batchelor's theory envisions \cite{davidson2015turbulence}. Therefore, a relevant approach to seek an answer to this question is to study the long-time behavior of enstrophy record that contains the spatially integrated information in the vortical motions over the entire domain and also is a representative for the dissipation dynamics. Similar to the kinetic energy \eqref{eqn: KE}, we define the enstrophy, $\mathcal{E}(t)$, in our problem setting as
\begin{align}
    \mathcal{E}(t) = \frac{1}{\mu(\pmb{\Omega})}\int_{\pmb{\Omega}} {\left\vert \omega^\prime_z(\mathbf{x},t) \right\vert}^2 d\pmb{\Omega}.
\end{align}
By computing the record of enstrophy for relatively long times (obtained from the same flow realization we studied its fluctuating velocity and vorticity behavior), studying the early-/long-time scaling trend of enstrophy would be possible. To perform this very study, the validity and stability of long-time evaluation of QoIs for stochastic mathematical models is of crucial importance to be considered and it has been addressed in multiple prior studies. For instance, Xiu and Karniadakis \cite{xiu2003modeling} used generalized polynomial chaos (gPC) with relatively high resolutions in order to study the long-time behavior of vorticity field for the flow past a cylinder under the uncertain inflow boundary conditions. In another study, Xiu and Hesthaven \cite{xiu2005high} employed high-order stochastic collocation methods to achieve stable second-order moment response to the stochastic differential equations at the long times. Moreover, Foo \textit{et al.} \cite{foo2008multi} utilized multi-element probabilistic collocation method (ME-PCM) with high resolution in random space to compute stable long-time flow records. Therefore, maintaining sufficiently high resolutions in discretization of random space is a key point. In our study, the high-resolution uni-variate PCM we employed to obtain the fluctuating flow fields (as described in section \ref{subsec: Fluctuations}) essentially guarantees the validity and statistical stability of our evaluations for the long-time fluctuating vortictiy field and computing the enstrophy record as illustrated in Figure \ref{fig: Scaling_enstrophy}. This plot shows that in terms of enstrophy time-scaling, we observe three stages of time. Here at stage (\textbf{I}), enstrophy behaves as $\mathcal{E}\sim t^{-2}$ (for $t < 2.5)$, however, after a transition period, stage (\textbf{II}), it persistently follows $\mathcal{E}\sim t^{-1/2}$ time-scaling in stage (\textbf{III}). At the third stage, this ``anomalous'' long-time scaling with $t^{-1/2}$ rather than the expected $t^{-1}$ scaling could essentially be interpreted as the result of an ``intensified'' mechanism for \textit{birth} and \textit{growth} of coherent vortices that live for effectively long periods of time during the evolution of this internal flow right after the occurrence of the flow instability. Figure \ref{fig: wz_coherent} portrays two snapshots of instantaneous vorticity field, $\omega_z$, on a segment of cylinder close to the wall to show the evolution and form of these coherent vortex structures survived the vortex filamentation process. We emphasize that the long life of the mature and relatively large coherent vortical zones (clearly visible and attached to the cylinder's wall) is the main reason of the anomalous enstrophy time-scaling we observe at stage (\textbf{III}) in Figure \ref{fig: Scaling_enstrophy}. As we mentioned earlier, this phenomenon could potentially be a practical engineering candidate to enhance and reinforce the effective chaotic/turbulent mixing qualities by inducing more memory effects resulted from a symmetry-breaking flow instability.

\section{Conclusion and Remarks}\label{sec: Conclusion}

The present study leverages the outcome of stochastic modeling and simulations to carry out a thorough analysis on the initiation of flow instabilities within high-speed rotating cylinders. Considering the random nature of the problem, a detailed mathematical representation of the stochastic incompressible Navier-Stokes equations was presented. Further, a high-fidelity stochastic CFD framework was introduced, which employs spectral/$hp$ element method in the forward solver and later on the stochastic space was numerically handled by probabilistic collocation method. Detailed grid generation steps and required convergence studies for the deterministic solver were obtained and stochastic discretization convergence were studied for the solutions of first and second moments. The time-evolution of expected kinetic energy of the flow in addition to its variance were computed and the uncertainty bounds propagated in the solution were identified with time. A variance-based sensitivity analysis of the random parameters of the model were conducted to globally characterize the most effective stochastic factor on the total variance of kinetic energy, consequently, the ``eccentric rotation'' was learned to be the dominant source of stochasticity. Later on, the expected solution from a very fine uni-variate PCM discretization on the dominant random parameter was utilized to compute the fluctuating velocity and vorticity fields for a randomly drawn realization of the sample space. These fluctuations were statistically analyzed through the time-evolution of their PDFs for radial and azimuth components in a qualitative manner while comparing to the standard Gaussian PDF. Statistical features such as appearance of intermittent and rare events in terms of heavy-tailed PDFs in addition to observing asymmetries in velocity and vorticity PDFs were spotted out. In particular, very close to the flow instability onset, these non-Gaussian statistical features were found to quickly get intensified especially for the radial velocity fluctuations and therefore fluctuating vorticity field as the flow evolves in time. Moreover, the statistics of flow fields were quantitatively measured through computing the skewness and flatness factors on narrow radial stripes extending from the wall to the cylinder's center. These records closely supported our qualitative findings from studying the PDFs of fluctuations and identified that in velocity field we quickly face regions with skewness factor of $\mathcal{O}(1)$ and flatness factor of $\mathcal{O}(10)$ while for the vorticity field these factors were recorded with about one order of magnitude higher than their velocity counterparts emphasizing on significantly high non-Gaussian vorticity induced by cylinder rotation affected by symmetry-breaking factors. Motivated by this observed strong non-Gaussianity, we sought to study the effects of coherent vortical structures essentially inducing memory effects into the vorticity dynamics. Thus, we managed to compute the time-scaling of the enstrophy record. Interestingly, we learned that unlike the early stages of flow after introduction symmetry-breaking rotational effects, enstrophy is scaled as $t^{-1/2}$ at long-time. This anomalous time-scaling essentially reveals the very existence of long-lasting and growing coherent vortical regions initially generated due to the non-symmetric rotation of the cylinder wall. This mechanism seems to be a promising engineering strategy to increase the chaotic/turbulent mixing time and quality for the rotating hydrodynamic systems.

\section*{Acknowledgement}

This work was supported by the MURI/ARO award (W911NF-15-1-0562), the AFOSR Young Investigator Program (YIP) award (FA9550-17-1-0150), the ARO YIP award (W911NF-19-1-0444), and the NSF award (DMS-1923201). The HPC resources and services were provided by the Institute for Cyber-Enabled Research (ICER) at Michigan State University. In addition, authors would like to thank Eduardo A. B. de Moraes for several helpful discussions on stochastic discretization and sensitivity analysis.  

\appendix
\section{Validation of Numerical Setup}\label{sec: Appendix1}

\begin{figure}[t!]
    \begin{minipage}[b]{.48\linewidth}
        \centering
        \includegraphics[width=1\textwidth]{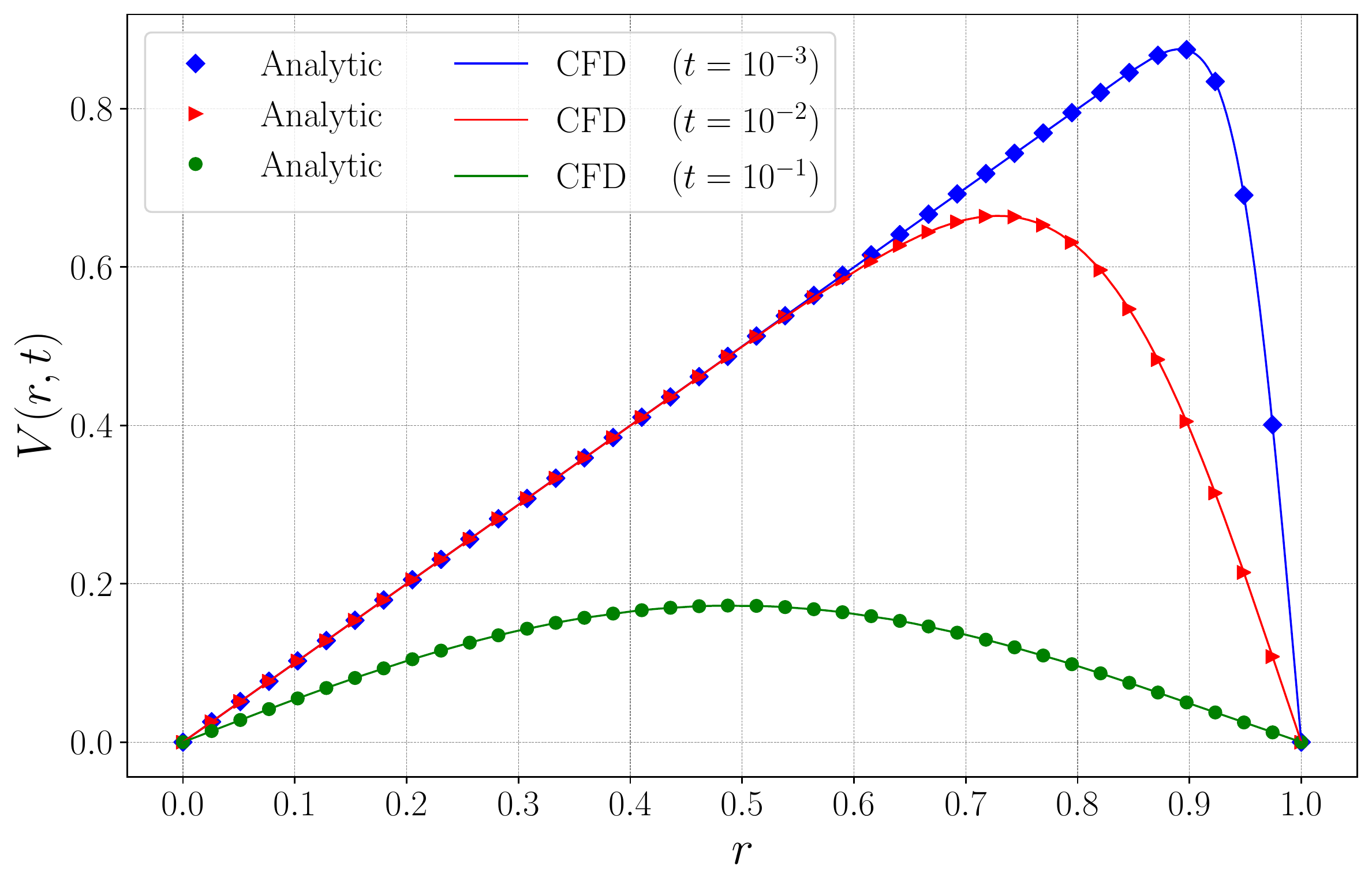}
        \subcaption{Complete sudden stop.}\label{fig: Comparison_a}
    \end{minipage}
    \begin{minipage}[b]{.02\linewidth}
        ~
    \end{minipage}
    \begin{minipage}[b]{.48\linewidth}
        \centering
        \includegraphics[width=1\textwidth]{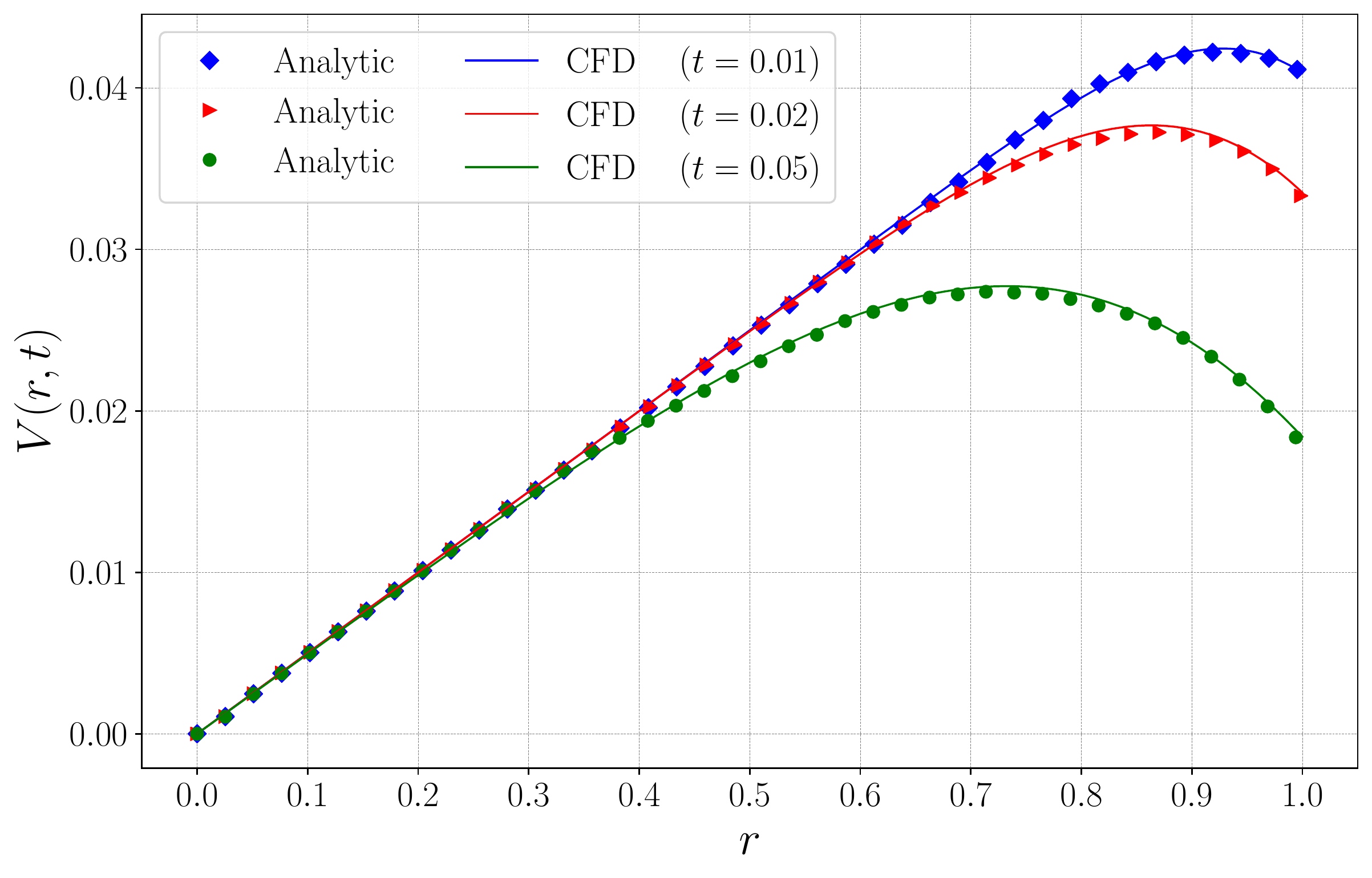}
        \subcaption{Exponential decay with $\frac{R_e}{R_{-}}=20$.}\label{fig: Comparison_b}
    \end{minipage}
    \caption{Comparison between the velocity, $V(r,t)$, obtained from CFD and analytical solution for flow at $Re=100$.}\label{fig: Comparison}
\end{figure}

This appendix provides a comparison study between the analytic and numerical solutions for specific cases of impulsive and exponential spin-decay at low-Reynolds numbers in order to validate our CFD results. Simplifying the governing equations in cylindrical coordinate system, $(r,\theta,z)$, for a non-stationary 2-D viscous incompressible flow, gives
\begin{align} \label{eqn: A1}
     &\rho \left(-\frac{u_\theta ^2}{r}\right) = - \frac{\partial p}{\partial r} ,\\ \nonumber
       &\rho \left( \frac{\partial u_\theta }{\partial t}\right) = \mu \bigg( \frac{\partial ^2 u_\theta}{\partial r^2}-\frac{1}{r^2} u_\theta + \frac{1}{r} \frac{\partial u_\theta}{\partial r} \bigg ).\nonumber
\end{align}
Here, the first and second equations represent the momentum equation in $r$ and $\theta$ directions, respectively.
By considering no-slip boundary conditions on the wall and taking the initial condition as $V(r,0)=r\dot{\theta}$ (rigid-body rotation), equation \eqref{eqn: A1} can be solved through the Laplace transform on the variable $t$ \cite{neitzel1980energy, kim2006onset}. If the length is scaled by the radius of cylinder, $r$, time is scaled by $r^2/\nu$, velocity in the \textit{sudden stop} case is scaled by $r\dot{\theta}$, and velocity in the \textit{exponential decay} case by $\lambda r^3 \dot{\theta} /\nu$, the resulting solution would be dimensionless. Therefore, the exact solutions for the complete sudden stop and exponential decay cases at low-Reynolds numbers are obtained as
\begin{align}\label{eqn: A2}
   & V_s(r,t)= -2 \sum_{n=1}^{\infty} \frac{J_1 (\beta_n r)}{\beta_n {J_0 (\beta_n)}} \exp(-\beta^2_n t),\\ \nonumber
    &V_e (r,t)= \frac{R_{-}}{R_{e}}\frac{J_1 (r \sqrt{B}) \exp(-Bt)}{J_1 (\sqrt{B})}+2 \sum_{n=1}^{\infty} \frac{J_1 (\beta_n r) \exp(-\beta^2_n t) }{\beta_n(\beta^2 - \beta) J_0(\beta_n)},
\end{align}
where $V_s(r,t)$ indicates the azimuth velocity for sudden stop case, $ V_e (r,t)$ is the azimuth velocity for the exponential decay case, $J$ is the Bessel function of the first kind, and $\beta_n$ denotes the positive roots of $J_1(\beta_n)=0$. Also $R_{-}=r^2 \dot{\theta}/\nu$ shows the Reynolds number corresponding to the initial state and $R_e=r^4\dot{\theta}\lambda / \nu^2$ denotes the Reynolds number for the spin-decay period (see [\onlinecite{neitzel1980energy}] and [\onlinecite{kim2006onset}] for derivations). Using equation \eqref{eqn: A2} and implementing the same initial and boundary conditions in the numerical setup for a low $Re$ number, a comparison in different times was made (see Figure \ref{fig: Comparison}). These comparisons are obtained for $Re=1/\nu=100$ and $R_e/R_{-}=20$, while we consider the mentioned dimensionless solution and the physical parameters. Comparing the analytic and the CFD results clearly validates our numerical implementation and procedure. It should be mentioned that the analytic solutions are only valid at the low-$Re$ number regime where no flow instability is created during these processes.

\section{Computational Workflow}\label{sec: Appendix2}

Performing numerous amount of forward simulations for discretization of random space urges the design of a proper workflow in high-performance computing (HPC) environment \cite{Callaghan2017rvgahp, Simakov2019}. In this work, we are dealing with a forward solver with requires input session files in the \texttt{xml} format, which contain information about the grid and each forward simulation's conditions. Using parallel computing on $\mathcal{O}(100)$ processes is inevitably demanded for each one of these forward simulations. Indeed, the number of simulations addressed in this work, could not be achieved by manually generation of input session files that are fed by realizations of stochastic parameter space. Hence, a Python program is prepared to construct the parameter space realizations (either from MC approach or PCM) and assign them to separate \texttt{xml} scripts that are placed in a directory associated with each forward simulation. Moreover, it enables automation of job submission step in the HPC environment. The statistical solutions (\textit{i.e.}, expected fields and their standard deviation) are computed by post-processing through \texttt{Paraview} toolkit. In particular, we exploit \texttt{Paraview}'s Python scripting (executed by \texttt{pvpython}) to extract the flow field variables from \texttt{xml} field files at the SEM integration points and perform required computations on them to obtain the expectation and standard deviation of field variables. Similar procedure is carried out to compute the velocity and vorticity fluctuation fields.


\bibliography{ref}

\end{document}